\documentclass[9pt,twocolumn,twoside,lineno]{pnas-new}
\usepackage{bm}
\usepackage{ulem}

\newcommand{\be}{\begin{equation}}
\newcommand{\ee}{\end{equation}}
\newcommand{\bea}{\begin{eqnarray}}
\newcommand{\eea}{\end{eqnarray}}

\templatetype{pnasresearcharticle} 

\title{Multi-scale structural complexity of natural patterns}

\author[a,b,1,2]{Andrey A. Bagrov}
\author[b,1]{Ilia A. Iakovlev}
\author[c]{Askar A. Iliasov}
\author[c,b]{Mikhail I. Katsnelson}
\author[b]{Vladimir V. Mazurenko}

\affil[a]{Department  of Physics  and  Astronomy,  Uppsala  University,  Box  516,  SE-75120  Uppsala, Sweden}
\affil[b]{Theoretical Physics and Applied Mathematics Department, Ural Federal University, 620002 Yekaterinburg, Russia}
\affil[c]{Institute for Molecules and Materials, Radboud University, Heyendaalseweg 135, 6525 AJ, Nijmegen, The Netherlands}

\leadauthor{Bagrov}

\significancestatement{Structural complexity of patterns, systems and processes is a very basic and intuitively clear concept in human’s perception of reality that is very difficult to describe quantitatively. A demand in a mathematical notion that properly reflects complexity of hierarchical non-random structures exists in many areas of science, from geology to social sciences. Here we propose an easy to compute, robust and universal definition of complexity based on inter-scale dissimilarity of patterns. Using classical magnetic patterns as an example, we demonstrate that our approach leads to maximization of complexity for the most visually non-trivial patterns, and can be used to detect phase transitions with high accuracy, making it a promising tool for studying pattern formation in a variety of systems.}

\authorcontributions{A.A.B. and M.I.K. suggested the idea and designed the project. I.A.I. performed simulations and produced the data. A.A.I. derived the analytical interpretation of structural complexity. A.A.B. wrote the manuscript with the help of V.V.M. and M.I.K. V.V.M. conducted the food dye experiments. A.A.B. and V.V.M. coordinated the project. All authors participated in the discussions. The authors declare no competing interests.}
\equalauthors{\textsuperscript{1}A.A.B. and I.A.I. contributed equally to this work.}
\correspondingauthor{\textsuperscript{2}andrey.bagrov@physics.uu.se}


\begin{abstract}
Complexity of patterns is a key information for human brain to differ objects of about the same size and shape. Like other innate human senses, the complexity perception cannot be easily quantified. We propose a transparent and universal machine method for estimating structural (effective) complexity of two- and three-dimensional patterns that can be straightforwardly generalized onto other classes of objects. It is based on multistep renormalization of the pattern of interest and computing the overlap between neighboring renormalized layers. This way, we can define a single number characterizing the structural complexity of an object. We apply this definition to quantify complexity of various magnetic patterns and demonstrate that not only does it reflect the intuitive feeling of what is ``complex'' and what is ``simple'', but also can be used to accurately detect different phase transitions and gain information about dynamics of non-equilibrium systems. When employed for that, the proposed scheme is much simpler and numerically cheaper than the standard methods based on computing correlation functions or using machine learning techniques. 
\end{abstract}

\dates{This manuscript was compiled on \today}
\doi{\url{www.pnas.org/cgi/doi/10.1073/pnas.XXXXXXXXXX}}

\begin{document}
\verticaladjustment{-2pt}

\maketitle
\thispagestyle{firststyle}
\ifthenelse{\boolean{shortarticle}}{\ifthenelse{\boolean{singlecolumn}}{\abscontentformatted}{\abscontent}}{}

\dropcap{C}omplexity is one of the most fundamental properties of the world around us and a key subject for many natural and social sciences; in  some of them, like biology, the origin of complexity is one of the central issues \cite{adami,gellmann,bakbook,badii,Koonin,KWK,WKK}. Despite numerous attempts to give a formal definition of complexity \cite{badii,MMC,GL,lloyd,CarrNetwork,Giulio,DeGiuli,beauty}, our understanding of these matters is still far from being complete. The famous motto ``I know it when I see it'' is definitely applicable to complexity but to formalize this feeling is a very nontrivial problem. One of the first and the most famous definitions, the Kolmogorov complexity, which is given in terms of the minimal instruction length required to describe the object \cite{Kolmogorov}, characterizes rather randomness and irregularity of the object than its structural non-triviality. Importantly, there is no general way to {\it calculate} the Kolmogorov complexity \cite{badii}. A different approach was taken by P. Bak and coauthors \cite{bakbook,bak1,bak2,bak3} who introduced the concept of self-organized criticality (SOC) as a universal root of structural complexity. Despite definite relevance of this concept to a number of natural and social phenomena such as \cite{geology, wars, biol}, and to the emergence of biological complexity \cite{WKK}, it does not give however a full satisfactory solution of the problem. Our intuitive perception of complexity is based on a tiny balance between how many different elements and connections the system has and how {\it recognizable} it is. The latter is usually related to having a reasonable number of distinguishable features at several well-separated characteristic scales. In other words, {\it complexity assumes hierarchy}. If we consider, for instance, ``complex'' structures in metallurgy, like pearlite colonies in steel \cite{RGK}, we deal with essentially different pictures at the atomistic scale within different phases (ferrite and cementite), at the scale of interphase boundaries, and at the scale of mesoscopic structure which is directly related  to mechanical properties. Coexistence of essentially different structural levels and competing constraints at these levels is also of crucial importance in biology \cite{WKK} and social sciences \cite{NonScaleFree}. This poses a natural question of how to account for this property quantitatively.

While there are many definitions of structural (or effective) complexity \cite{GL,Bennett,Crutchfield}, most of them have a common weakness: in each particular case, one must decide subjectively what is essential structural features, and what is mere a noise which must be ignored. In principle, there is nothing wrong with complexity being context-dependent and a bit subjective. Still, it is tempting to find a way to define complexity as a more ``observer-independent'' quantity that can be used in different contexts with only slight modifications. With this in mind, a natural list of requirements for a proper notion of structural complexity can be formulated:
\begin{itemize}
    \item It must aggregate information about different scales present in the problem. 
    \item It must be well-defined analytically, so that for the selected class of objects (patterns, texts, melodies etc.) the protocol of computing it can be executed with a little need to make subjective choices and decisions.
    \item Within the same class of objects, it must be robust and stable upon reasonably mild deformations of an object.
    \item It should be small both for trivially ordered and fully disordered structures.
\end{itemize}

Among other things, a promising view on these matters was formulated in \cite{Wolpert1, Wolpert2}, where, for a broad variety of structures, a clearly defined and computable measure of {\it{ self-dissimilarity complexity}} was given (see also \cite{Lloyd-depth} for a similar in spirit approach). It was suggested that a structure is the more complex the more it differs from itself when considered at different spatial and temporal scales. 

The idea of relating complexity of a pattern or a structure to a certain functional over all scales has also been discussed more pragmatically in concrete physical contexts. For example, in the theory of polymers, it was suggested to study conformational properties of proteins by analyzing how certain observables scale upon Renormalization Group (RG) transformations and keeping track of the whole RG flow profile, not only the deep infrared behavior \cite{Sinelnikova}. Another research area where the concept of complexity has attracted considerable attention is the anti de Sitter/Conformal Field Theory correspondence (also known as holography). There it was conjectured that computational complexity of a quantum state should be related to the volume of dual bulk space which, in holographic terms, means integration over all the involved energy scales \cite{Susskind:2014, Brown:2015}, and possible conceptual connections to SOC were discussed \cite{Bagrov_complexity}.

Inspired by these attempts, we give a quantitative definition of structural complexity of patterns in terms of RG flow. A pattern can be regarded as a function $f(x)$ defined on a certain domain $D$. For example, a gray-scale picture is a real-valued function on a two-dimensional rectangle. For such an object, RG transformation can be defined in a natural way. E.g., if $D$ is a discrete set of pixels or lattice sites, a coarse-grained pattern can be obtained by means of Kadanoff decimation. If it is a continuous domain, RG transformation can be implemented as convolution of $f(x)$ with some scale-dependent filter. It is natural to say that scale $\lambda$ contributes some features to the pattern if there is a difference between the coarse-grained pattern $f_\lambda(x)$ and its a bit more coarser version $f_{\lambda+d\lambda}(x)$\footnote{This notation assumes that, if $L$ is the linear size of the pattern and $\Lambda$ is the filter width, $f_L(x)$ is the original pattern before renormalization, and $f_\Lambda(x)$ is the most coarse-grained version of it}. The latter can be measured as deviation \begin{gather}
   {\cal C}_\lambda= |\langle f_\lambda(x)|f_{\lambda+d\lambda}(x)\rangle -\\ \frac12\left( \langle f_\lambda(x)|f_\lambda(x)\rangle + \langle f_{\lambda+d\lambda}(x)|f_{\lambda+d\lambda}(x)\rangle\right)|=\nonumber \\
   \frac{1}{2} \int_D \left( f_{\lambda+d\lambda}(x) -f_{\lambda}(x) \right)^2 dx,
   \nonumber
\end{gather} where $\langle f(x)| g(x) \rangle = \int_D dx f(x) g(x)$ is a non-normalized overlap of two patterns\footnote{Independently normalizing each scale to the same number can be a cause of undesired artifacts. For example, it might happen that the overall ``intenisty'' of a picture is decreased in the coarse-graining procedure, and in that case we want to regard large-scale contributions to the overall complexity value as small. Hence normalization of each scale will make the less intense patterns more important than they should be.}. Summing up this over all scales, we obtain a number that we call {\it multi-scale structural complexity} $\cal C$:
\begin{equation}\label{eq:C_definition}
    {\cal C} = \sum\limits_\lambda {\cal C}_\lambda.
\end{equation}

While this approach is quite generic and allows to estimate complexity of almost any structure for which the coarse-graining procedure can be defined, here we focus on several concrete examples to demonstrate how the concept of structural complexity can be utilized to address physical problems. 
First, we study the phase transitions in the $2d$ and $3d$ classical Ising model, and demonstrate that complexity of the critical point is indeed higher than that of the fully ordered ferromagnetic phase or fully random paramagnetic one. Moreover, we show that one can compute $T_c$ with high accuracy simply by looking at the temperature dependence of complexity. For each value of $T$ it is enough to compute ${\cal C}(T)$ just for a single snapshot of the system, without any need to compute correlation functions and average over multiple Monte Carlo samples.

From that we proceed to a more complicated classical Heisenberg model with Dzyaloshinskii–Moriya interactions which hosts a variety of phases that cannot be characterized with a local order parameter but appear to be non-trivial patterns, such as spin spirals, bimerons, and skyrmion crystals. Again, not only the suggested multi-scale structural complexity maximizes on the most visually non-trivial spin spirals (magnetic labyrinths) and minimizes on the ordered ferromagnetic configurations, but transition lines between the phases can be easily determined by computing complexity of mere single realizations of the spin configuration at each point of the phase diagram.

Finally, we study evolution of complexity in two time-dependent settings. As a warm-up, we consider a dye drop dissolving in water. This is an archetypal example of a process where entropy of a system grows steadily, but the apparent structural complexity evolves in a non-monotonous way. Computing $\cal C$ of snapshots of this process made at different moments of time, we show that the multi-scale complexity we defined indeed attains its maximum not on the most random configurations and demonstrates very appealing robustness of the pattern of temporal evolution for different runs of the experiment. Then we perform simulations of real-time spin dynamics in Dzyaloshinskii-Moriya ferromagnet and derive complexity of several non-equilibrium processes such as switching and breathing of skyrmions \cite{Blugel}, and melting of magnetic labyrinths. We show that in all these cases evolution of complexity properly reflects the spin dynamics, and is fully in line with intuitive expectations.

We complement our numerical studies with an analytical view on the suggested notion of structural complexity and demonstrate that, in the case of discrete renormalization group protocol, complexity of a pattern can be formulated as a Jackson integral from quantum calculus, and briefly discuss possible roots of this, at first glance unexpected, connection.
\subsection*{Method}
\begin{figure}[!h]
\begin{center}
\includegraphics[width=\columnwidth]{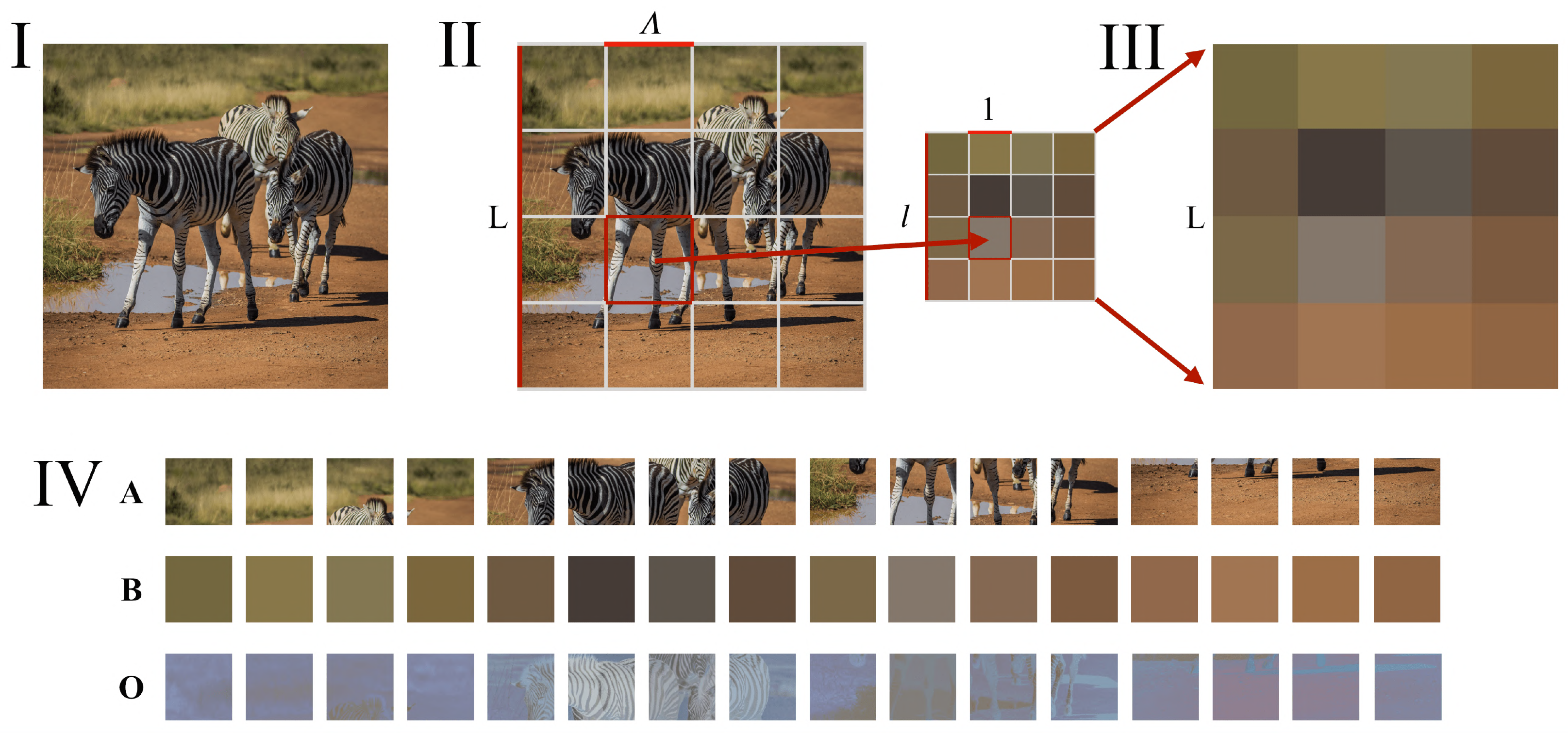}
\end{center}
\caption{Schematic representation of the idea behind the proposed method. A photo of $L \times L$ pixels (panel I) taken from www.pexels.com is divided into blocks of $\Lambda \times \Lambda$ pixels (panel II). A renormalized photo of $l \times l$ pixels is plotted, where $l = L / \Lambda$  ($l=4$, $\Lambda=2$ in this example). The renormalized photo is rescaled up to the initial photo size (panel III). Vectors ${\bf A}$ and ${\bf B}$ are constructed from blocks of the initial and the renormalized images respectively (panel IV). The scalar product of these vectors is used to define overlap $O$. For illustrative purposes, pixelwise products of $\bf A$- and $\bf B$-blocks  are shown as vector $\bf O$. Corresponding complexity obtained with $N_o$ = 10 and $\Lambda$ = 2 is equal to ${\cal C} = 0.163859$}
\label{fig1}
\end{figure}

To demonstrate how complexity of a pattern can be computed, we shall consider a photo of $L \times L$ pixels as an example \cite{pexels}, Fig.\ref{fig1}. Position of each pixel is given by its row and column indices $i,j$, and its state is characterized in general case by some vector ${\bf s}_{ij}$. The meaning of the vector depends on the context. For example, in the case of a color picture, it is a three-dimensional vector which components encode the color in the RGB scheme scaled to the range [-1; 1]. For a magnetic system, they will be the $x$, $y$, and $z$ components of the spin. In the simpler case of gray-scale pictures or magnetic patterns characterized only by $z$-projection of local magnetization, state of a pixel will be a single number instead.

The original pattern is then renormalized (coarse-grained) in a certain way. There could be different approaches to renormalization, -- the picture can be convolved with e.g. a Gaussian filter, or some more sophisticated scheme can be implemented, like the one defined in \cite{Sinelnikova} for polymer chains. Obviously, the resulting value of complexity will be dependent on the employed scheme. However, we found that already the simplest discrete decimation scheme leads to meaningful and robust results, so in the rest of the paper we stick to it\footnote{A similar approach has been taken in \cite{Bialek1, Bialek2}, where coarse-graining has been used to demonstrate scale-invariant features of natural landscapes.}. 

At each iteration, the whole system is divided into blocks of $\Lambda \times \Lambda$ size, and each block is substituted with a single pixel which state is calculated as ${\bf s}_{ij} (k) = \frac{1}{\Lambda^2} \sum_{l} \sum_{m} {\bf s}_{\Lambda i+m, \Lambda j+l} (k-1)$, where the $lm$ indices enumerate the pixels belonging to the same block, and $k$ is the number of iteration. This procedure is then repeated several times resulting in a stack of renormalized patterns of different resolution. With such a stack at hands, we can compute overlaps between patterns separated by one step of renormalization group. To do that, in every pair, the ``coarser'' pattern is rescaled up to the linear size of the ``finer'' one to keep the number of pixels in them the same. A schematic visualization of this is given in Fig.\ref{fig1}.:
\begin{gather}
O_{k, k-1} = \frac{1}{L_{k-1}^2} \sum_{i=1}^{L_{k}} \sum_{j=1}^{L_{k}}  {\bf s}_{ij} (k) \cdot \sum_{m=1}^{\Lambda} \sum_{l=1}^{\Lambda} {\bf s}_{\Lambda i+m, \Lambda j+l} (k-1) \\ =\frac{\Lambda^2}{L^2_{k-1}}\sum_{i=1}^{L_{k}} \sum_{j=1}^{L_{k}}  {\bf s}^2_{ij} (k) = \frac{\Lambda^2}{L^2_{k-1}} \cdot L^2_{k}\cdot O_{k,k}=O_{k,k}, \nonumber
\end{gather}
with $k=0$ corresponding to the original pattern, and $O_{k,k}$ is an overlap of the pattern at scale $k$ with its own self. Note that the overlap defined this way is not normalized, $O_{k,k}\not\equiv 1$, and $O_{k,k-1}= O_{k,k}$ only for the particular decimation scheme based on averaging, while being more non-trivial for other types of renormalization.
Defining structural complexity $\cal C$ as an integral characteristic accounting for features emerging at every new scale, we obtain
\begin{gather}
\label{eq:Complexity}
{\cal C} = \sum\limits_{k=0}^{N-1} {\cal C}_k =  \sum\limits_{k=0}^{N-1} |O_{k+1,k} - \frac12\left(O_{k, k}+O_{k+1, k+1}\right)|,
\end{gather}
where $N$ is the total number of renormalization steps.

\subsection*{Complexity of artificial and natural structures}

\begin{figure}[!h]
\begin{center}
\includegraphics[width=\columnwidth]{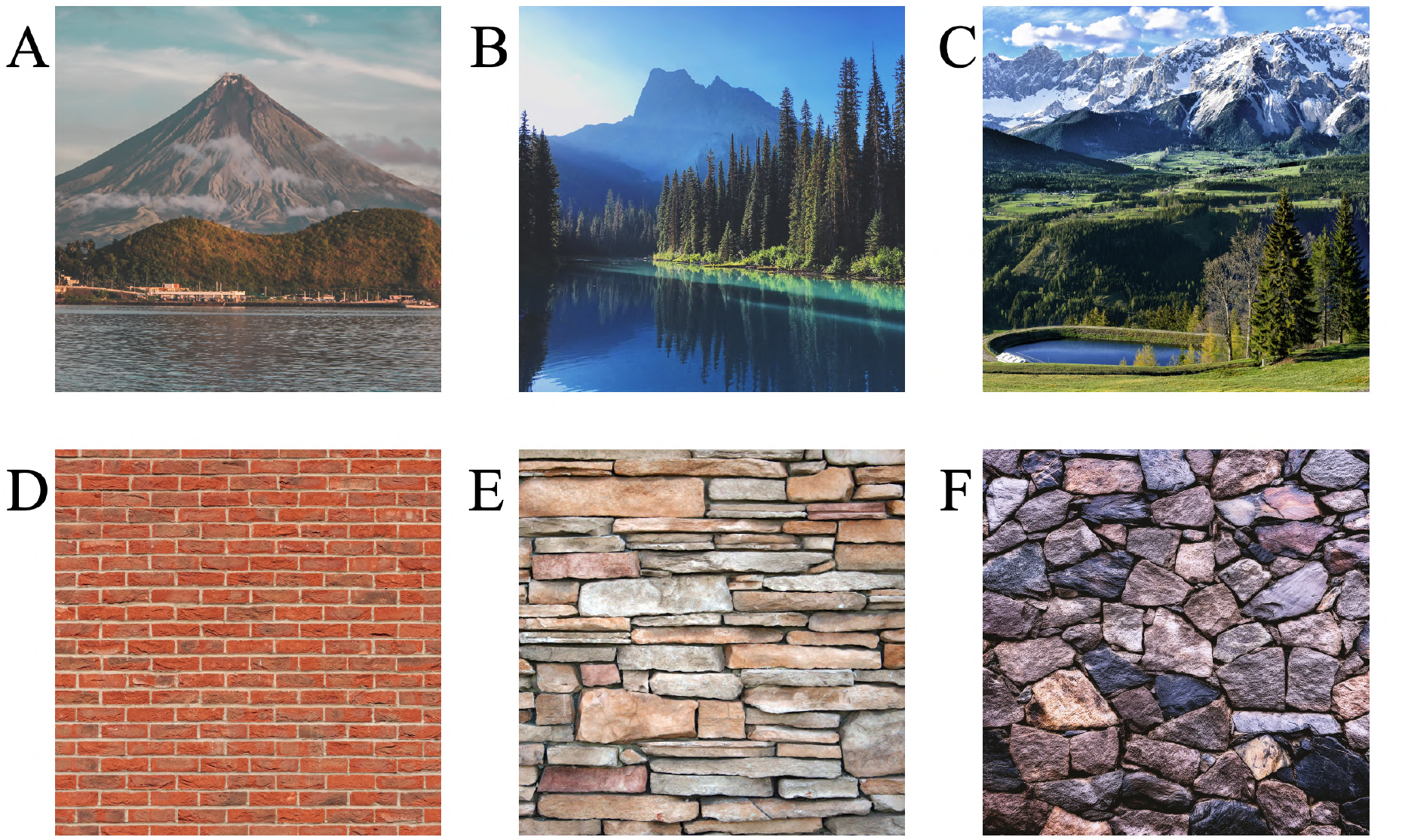}
\end{center}
\caption{Structural complexity of natural (A-C) and artificial (D-F) patterns. The images of 4096$\times$4096 pixels were taken from \cite{pexels}. Corresponding complexities are equal to (A) ${\cal C} = 0.078648$, (B) ${\cal C} = 0.135672$, (C) ${\cal C} = 0.272874$, (D) ${\cal C} = 0.107577$, (E) ${\cal C} = 0.276524$ and (F) ${\cal C} = 0.497536$. Here we used $N_o$ = 10 and $\Lambda$ = 2.}
\label{walls}
\end{figure}

As the first illustrative application of the developed method for estimating the structural complexity, we have chosen photos of different natural landscapes and walls. The former serve as examples of images, and the latter - as examples of more homogeneous textures. These images were downloaded in high resolution of 4096$\times$4096 pixels \cite{pexels}, which allows to perform up to 10 renormalization steps. It is clear that to build a wall using stones of different shapes and sizes is a more challenging task than to build a regular brick wall, and that a landscape combining short- and long-range features appears visually more complex than the one composed of only a few large elements. One can see from Fig.\ref{walls} that our numerical estimation of the complexity fully reflects these observations.

\subsection*{Phase transitions in Ising model}To elaborate on how the measure of multi-scale complexity can be employed to help answer concrete questions arising in different areas of physics, we first focus on one particular example -- the problem of constructing phase diagrams of statistical systems (see \cite{CarrNetwork, Danilov} for the complexity view on transitions in quantum systems). Even when order parameter is known, to determine the transition lines in the space of parameters might require extensive Monte Carlo simulations. The situation becomes much trickier if the order parameter is unknown, or if the transition is of unconventional nature (e.g. topological phase transitions).

Recently, an automatic way to detecting phase boundaries based on machine learning methods has been suggested \cite{Melko,Iakovlev1}. Since a neural network is dealing directly with {\it patterns}, the success of this approach poses a natural questions whether states of a system belonging to different phases can be distinguished by calculating their structural complexity.

To check this, we first consider the classical Ising model with nearest-neighbor ferromagnetic exchange interaction on square ($2d$) and cubic ($3d$) lattices:
\begin{equation}
H = -J \sum_{nn'} S^z_n S^z_{n'},\,\,J>0, \label{eq:Ising}
\end{equation}
and consider the paramagnetic/ferromagnetic phase transitions both in two and three dimensions. Then we study how complexity changes across the transition point. 

In the $2d$ case, we perform classical Monte Carlo simulations for \eqref{eq:Ising} on square lattice of 1024$\times $1024 size scanning over temperatures $0<T/J<4.5$ with step $\Delta T = 0.045 J$. For a lattice of this size, one can do eight renormalization steps within the proposed scheme. In $3d$, we conduct the same analysis for the Ising model defined on cubic lattice of 256$\times$256$\times$256 spins with the smallest possible 2$\times$2$\times$2 renormalization block, and scanning over $2<T/J<6.5$, $\Delta T = 0.045 J$.

\begin{figure}[!h]
\begin{center}
\includegraphics[width=0.8\columnwidth]{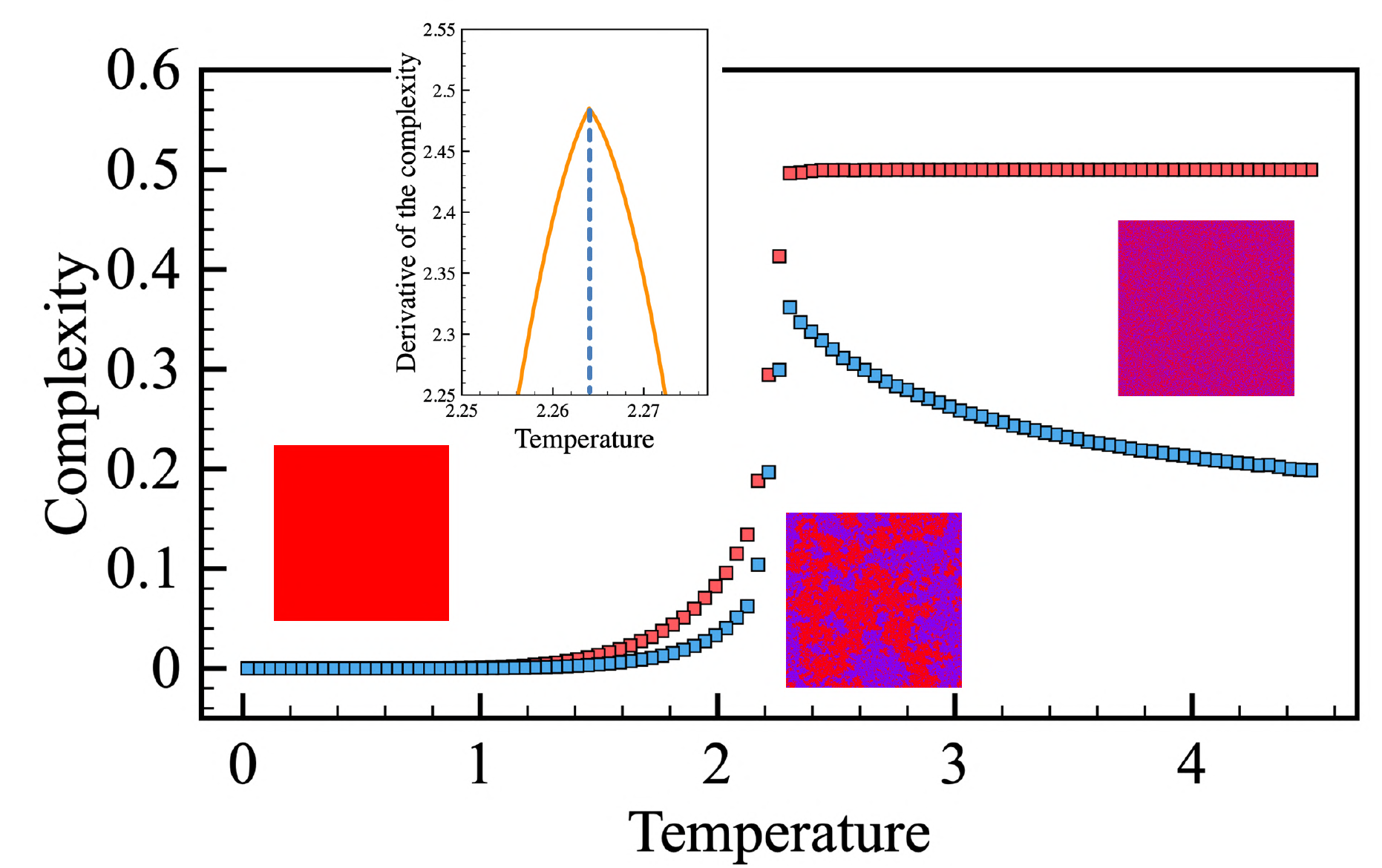}
\end{center}
\caption{Temperature dependence of the complexity obtained from the two-dimensional Ising model simulations. Red and blue squares correspond to the complexities calculated with $k \ge 0$ and $k \ge 1$, respectively. The size of error bars is smaller than the symbol size. Inset shows the first derivative of the complexity used for accurate detection of the critical temperature. Here we used $N$ = 8,  $\Lambda$ = 2.}
\label{ising_2d}
\end{figure}

\begin{figure}[!h]
\begin{center}
\includegraphics[width=0.8\columnwidth]{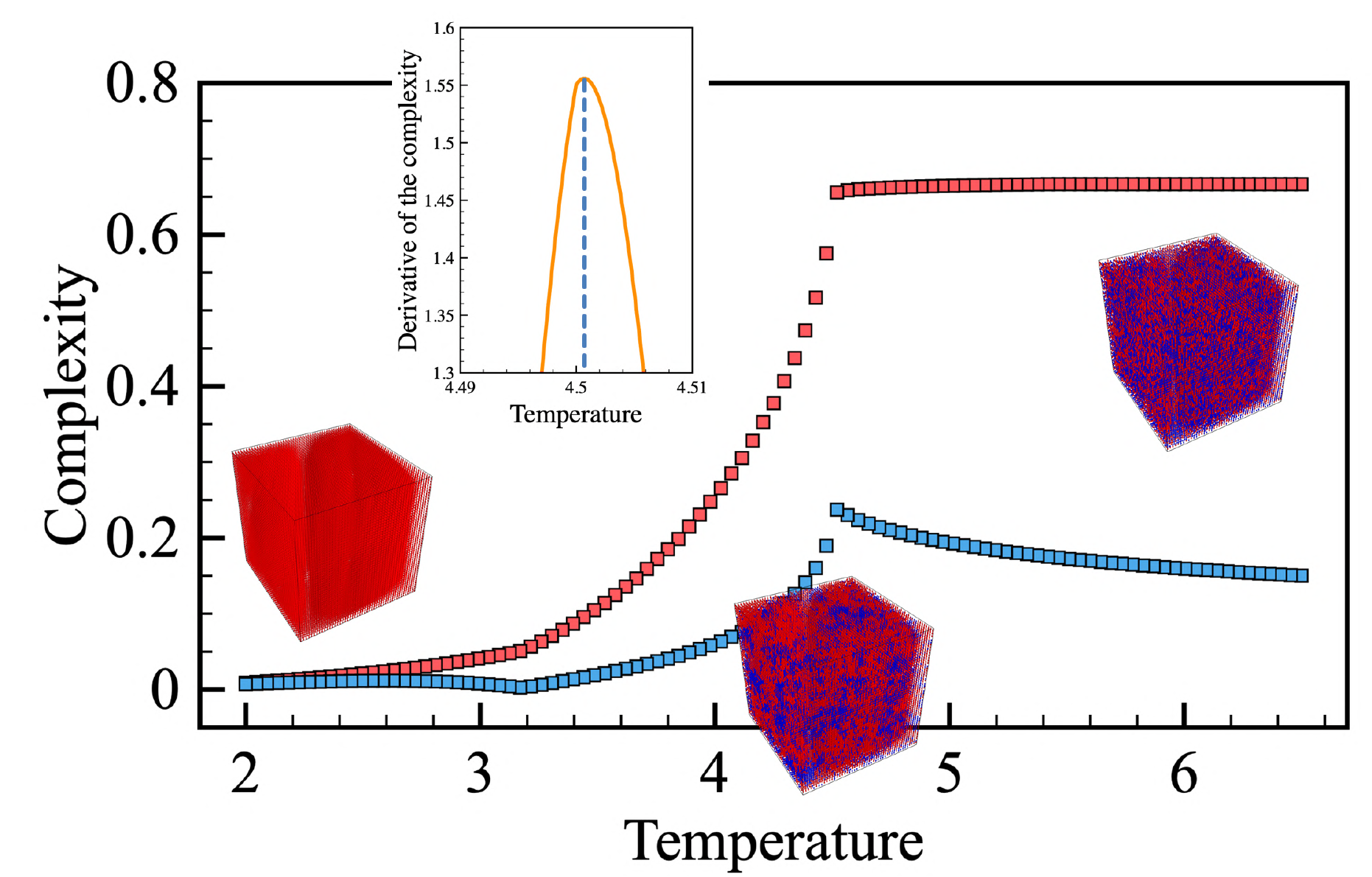}
\end{center}
\caption{Temperature dependence of the complexity obtained from the three-dimensional Ising model simulations with $\Lambda = 2$. Red and blue squares correspond to the complexities calculated with $k \ge 0$ and $k \ge 1$, respectively. The size of error bars is smaller than the symbol size. Inset shows the first derivative of the complexity used for accurate detection of the critical temperature. Here we used $L\times L\times L$ cubic lattice with $L = 256$, $N = 6$. The small but visible cusp on the blue curve around $T\simeq3.2$ reflects the emergence of magnetic domains within the ferromagnetic phase, which takes place sometimes during MC simulations on large lattices.}
\label{ising_3d}
\end{figure}

Structural complexity as a function of temperature is presented in Figs.~\ref{ising_2d},~\ref{ising_3d}. First thing interesting to note is that structural complexity of the Ising lattice configurations is very robust. Both in $2d$ and $3d$, for each value of $T$ we generated five different Monte Carlo samples, and their complexity turned out to be the same with very high accuracy (about $\sim 0.01\%$), thus we do not even show the error bars on the plot.

One can see that by taking derivative with respect to $T$ and associating the phase transition with the extremum of $d{\cal C}/dT$, the critical temperature can be estimated with very high accuracy. For the square lattice, our approach gives the value of $T/J\approx2.26$, which is in excellent agreement with known analytical results \cite{Onsager} $T_{c} / J = 2/ ln (1+\sqrt{2})\approx 2.269$. For the cubic one, we obtain $T_c \approx 4.5$, which is very close to the results of the high-temperature series expansion $T_c\approx 4.5103$ \cite{Fisher} and Monte Carlo simulations $T_c\approx 4.5$ \cite{MC}. Note that sometimes MC simulations lead to metastable configurations of magnetic domains inserted into the ferromagnetic phase, - and the structural complexity keeps track of that as well, Fig.~\ref{ising_3d}.

A peculiar detail of the ${\cal C}(T)$ dependence is that it saturates and reaches a constant value in the paramagnetic phase. This seemingly contradicts our intention to define {\it structural} complexity because the magnetization patterns at $T>T_c$ look visually more random and less structured than the critical point. However, two aspects should be kept in mind. First, if we neglect the contribution of the most microscopic scale that can be barely resolved visually (i.e. the $|O_{0,1}-\frac12(O_{0,0}+O_{1,1})|$ term in \eqref{eq:Complexity}), the resulting complexity of paramagnet would be smaller than that of the critical point, and will be decreasing with temperature. This fact speaks in a favor of the suggested definition as it is natural to expect structural complexity to depend on the resolution of a perceiver (be it a human being, a detector, or a neural network). Secondly, as we will discuss in more detail later on, apart from the single numerical value $\cal C$, another important property of a structure is how complexity is {\it distributed} between different scales. In the case of paramagnet, it comes mainly from the finest scale $k=0$, while for the more non-trivial structures it resides on a number of scales.

\subsection*{Complexity of spin textures}Our next goal is to see if the notion of structural complexity can be employed to detect phase transitions of a more sophisticated nature. An illustrative example of a system where complex patterns emerge naturally is magnets with Dzyaloshinskii-Moriya (DM) interactions described by the Hamiltonian \cite{Dzyaloshinskii, Moriya}: \begin{equation}\label{Ham}
H=- J\sum_{nn'}{\bf S}_n{\bf S}_{n'}-{\bf D}\sum_{nn'}[{\bf S}_n\times{\bf S}_{n'}]-  \sum_n B S_n^z,
\end{equation}
where $J$ and $\bf D$ are the isotropic exchange and DM interactions respectively, and the sums run over links of two-dimensional square lattice. Vector $\bf D$ is orthogonal to the lattice links.

Depending on the relative strength of interactions and the magnetic field, the magnet exhibits clearly distinguishable textures such as spin spirals, skyrmion crystals, and bimerons. Contra to the case of ferromagnetic/paramagnetic phase transition, transition between two types of textures cannot be related to symmetry breaking and described in terms of local order parameter. At the same time, it is clearly a physical effect that should be amenable to quantification. In our analysis, we consider a square lattice of 1024$\times$1024 size with $J=1,\, |{\bf D}|=1$, and perform Monte Carlo simulations at fixed temperature $T=0.02$ varying the external magnetic field B in the range $0<B<1$ with step $\Delta B=0.01$. For each value of B, we assume that the state of a lattice site (``pixel'' of the corresponding pattern) is characterized by components of spin.

\begin{figure}[!h]
\begin{center}
\includegraphics[width=\columnwidth]{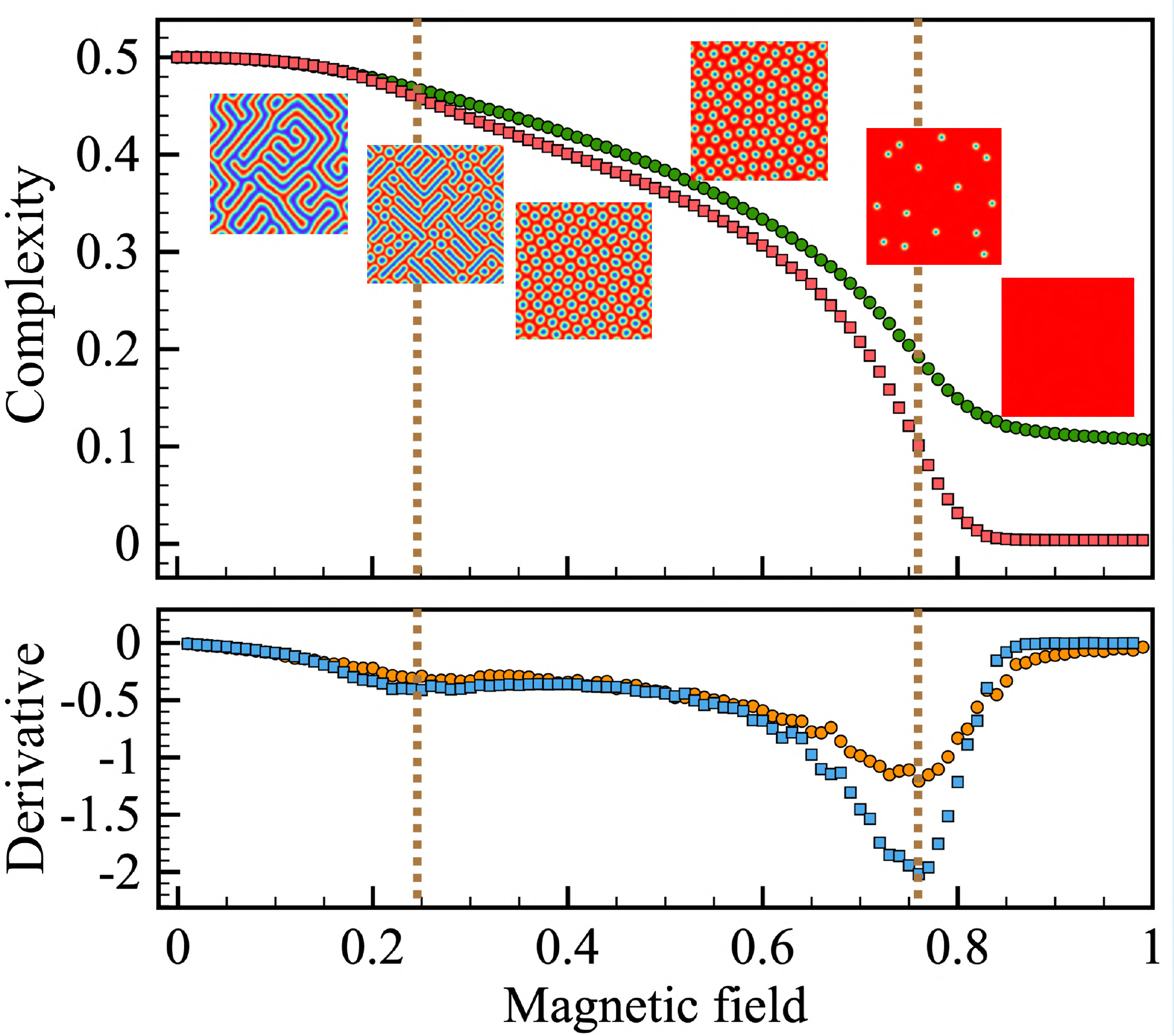}
\end{center}
\caption{(a) Magnetic field dependence of the complexity obtained from the simulations with spin Hamiltonian containing DM interaction with $J$  = 1 and $|{\bf D}|$ = 1. The error bars are smaller than the symbol size. (b) Complexity derivative we used for accurate detection of the phases boundaries. Squares and circles correspond to the simulations carried out at $T$ = 0.02 and $T$ = 0.4, respectively.}
\label{skyrmion}
\end{figure}

\begin{figure}[!h]
\begin{center}
\includegraphics[width=\columnwidth]{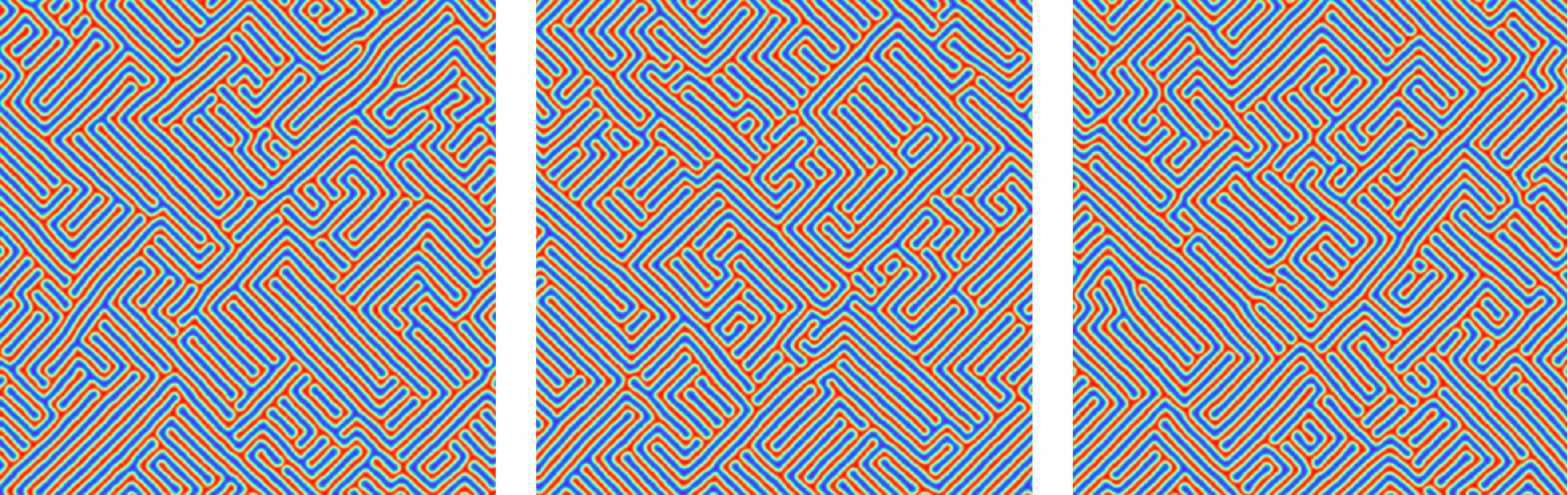}
\end{center}
\caption{Configurations of the DM magnetic on $1024\times 1024$ square lattice obtained from independent Monte Carlo runs with parameters $B=0.05 J$, $|{\bf D}|=J$, $T=0.02J$. While they are visually distinct, corresponding complexities (left to right) are equal to ${\cal C} = 0.4992115$, ${\cal C} = 0.4991825$ and ${\cal C} = 0.4991805$.}
\label{fig:robustness}
\end{figure}

The resulting dependence of complexity on magnetic field is presented in Fig.\ref{skyrmion}. Again, for each value of $B$ the complexity appears to be very robust, fluctuating within $0.01\%$ error range for independent Monte Carlo runs, Fig.~\ref{fig:robustness}. As before for the paramagnetic/ferromagnetic phase transitions, the extrema of complexity derivatives $d{\cal C}/dB$ reflect very well both the melting of spin spirals (magnetic labyrinths) into skyrmion crystals, with the transition point being exactly the bimeron phase, as well as the transition between skyrmion crystals and ferromagnets. 

An intriguing feature of ${\cal C}(B)$ is that the visually most complex magnetic configurations of labyrinth type that emerge at weak magnetic fields have the largest $\cal C$ value, which is yet another argument in favor of the interscale approach to defining effective complexity. Transitions between spin textures in DM magnets are a clear example of truly non-trivial physical application of structural complexity. Formation of a skyrmion crystal from decaying spin spiral cannot be detected with conventional observables, such as as magnetization and skyrmion number. Of course, it can be identified by a trained neural network \cite{Iakovlev1} but it would require learning a network on a large set of configurations. Instead, thanks to the robustness of $\cal C$ upon choosing different patterns at the same point of parametric space, one can resort to computing complexity of a {\emph{single}} Monte Carlo sample for each value of $B$ and find the transition point with much lesser effort.

It is also instructive to show that in some cases structural complexity even allows to detect phase transitions that cannot be captured by looking at the correlation functions of the state averaged over a number of configurations. Recently, some of us have shown~\cite{Iakovlev2} that all the phases of the DM ferromagnet can be identified at low temperature by calculating the spin structure factor averaged over a number of Monte Carlo samples at each point of the phase diagram:
\begin{equation}\label{hi2}
\chi_\perp(\textbf{q})=\frac{1}{N}\left\langle\left|\sum_{n}S^x_ne^{-i\textbf{q}\textbf{r}_n}\right|^2+\left|\sum_{i}S^y_ne^{-i\textbf{q}\textbf{r}_n}\right|^2\right\rangle,
\end{equation}
where $\textbf{q}$ is the reciprocal space vector, $S_n^\alpha$ ($\alpha = (x,y)$) is the projection of the $n$th spin and $\textbf{r}_n$ is the radius vector for the $n$th site.

\begin{figure}[!h]
\begin{center}
\includegraphics[width=\columnwidth]{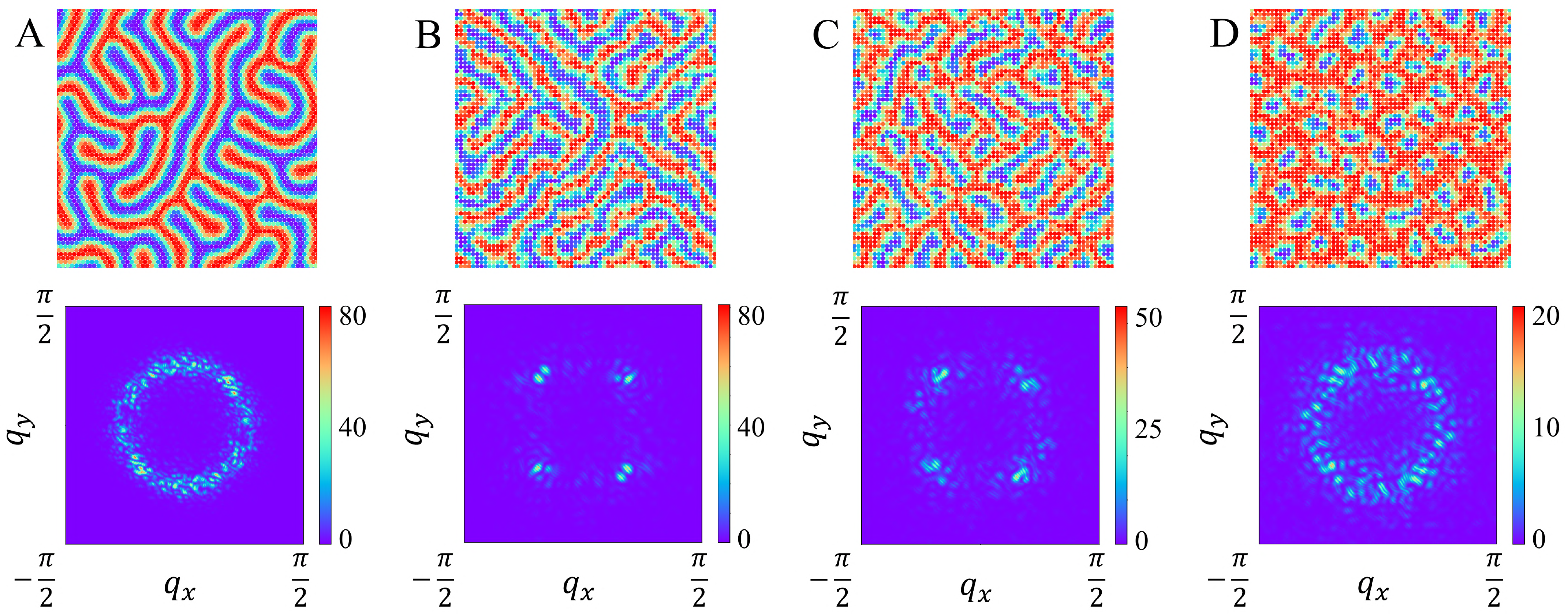}
\end{center}
\caption{(A) Magnetic labyrinth on a triangular lattice at low temperature ($T=0.02J$), (B) spin spirals, (C) mixed skyrmion-bimeron magnetic configuration and (D) pure skyrmions on a square lattice at high temperature ($T=0.4J$), and the corresponding spin structure factors. The complexities are equal to (left to right) ${\cal C} = 0.499879$, ${\cal C} = 0.500003$, ${\cal C} = 0.479131$ and ${\cal C} = 0.421104$.}
\label{struct}
\end{figure}

However, as it was shown in~\cite{Iakovlev1}, this method cannot be applied to the square-lattice DM ferromagnet at temperatures above $T\simeq0.3J-0.4J$. Moreover, in the case of triangular lattice, it fails to distinguish between spin spiral and skyrmion crystal phases even at low temperatures $T\simeq 0.02J$. The structure factors are then smeared out, Fig.~\ref{struct}, and the phase boundaries in the parametric space cannot be properly identified. At the same time, as can be seen from Fig.~\ref{skyrmion}, structural complexity as a function of magnetic field at $T=0.4J$ behaves similarly to the low temperature case allowing to detect the phase transitions.

\subsection*{Interscale distributions of complexity} 
\begin{figure}[!h]
\begin{center}
\includegraphics[width=\columnwidth]{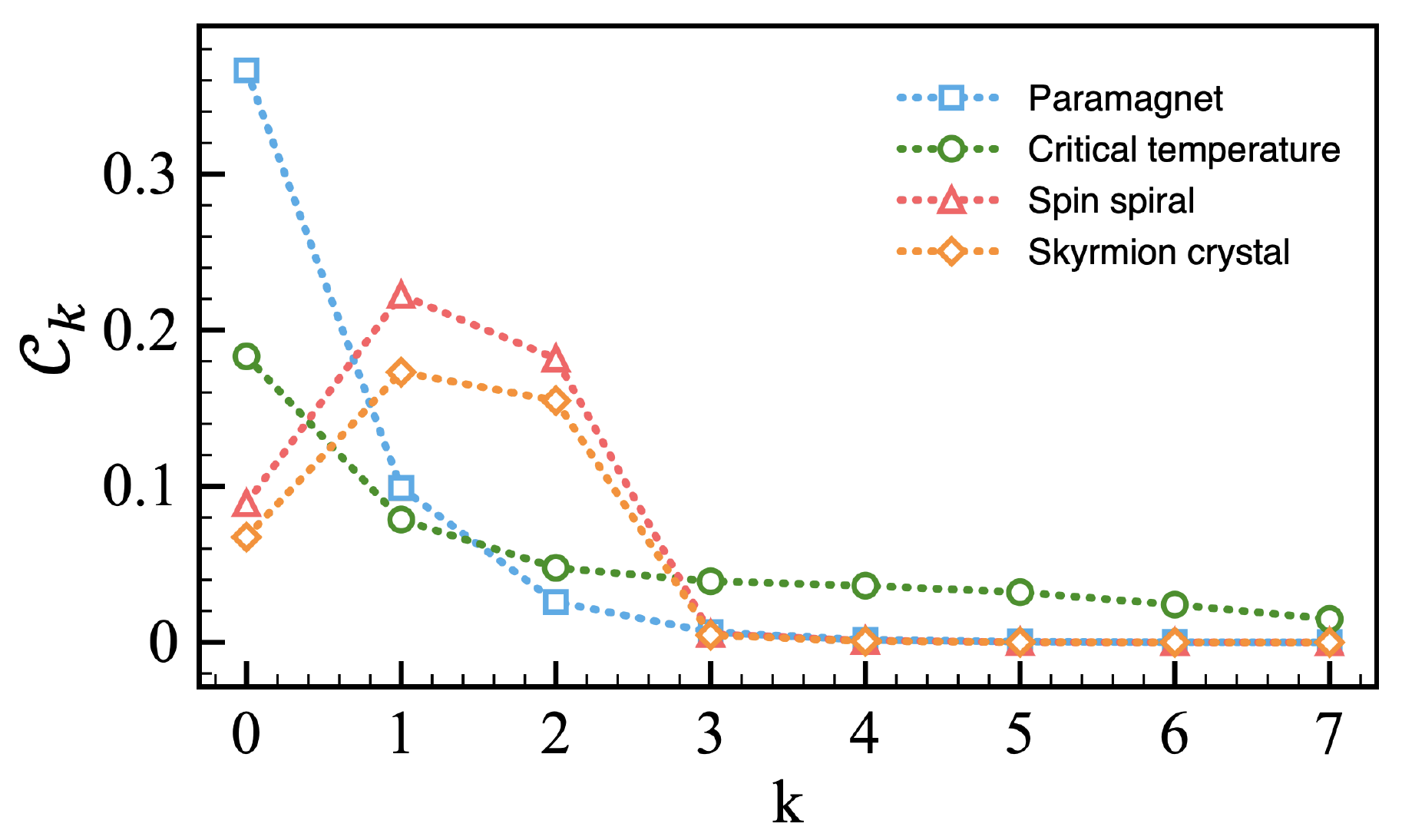}
\end{center}
\caption{Partial contributions of different scales to the overall structural complexity for four types of magnetic patterns.}
\label{fig:partials}
\end{figure}

As we briefly mentioned before, the absolute value of complexity $\cal C$ is not the only interesting quantity. More can be learned from how different scales contribute to structural complexity of a pattern. Thus, it is instructive to look at the scale distribution of partial complexities ${\cal C}_k$ for the four studied types of $2d$ patterns - spin spirals, skyrmion crystals, Ising spins at criticality and paramagnets. Those are plotted in Fig. \ref{fig:partials}. One can see that the most visually non-trivial configurations (spirals and crystals) are characterized by a couple of scales, complexity of the critical point is distributed pretty homogeneously among all the involved scales, which is what one would expect for a scale-invariant system, and complexity of a random paramagnetic pattern is strongly dominated by its deep microscopics.

\begin{figure}[!h]
\begin{center}
\includegraphics[width=\columnwidth]{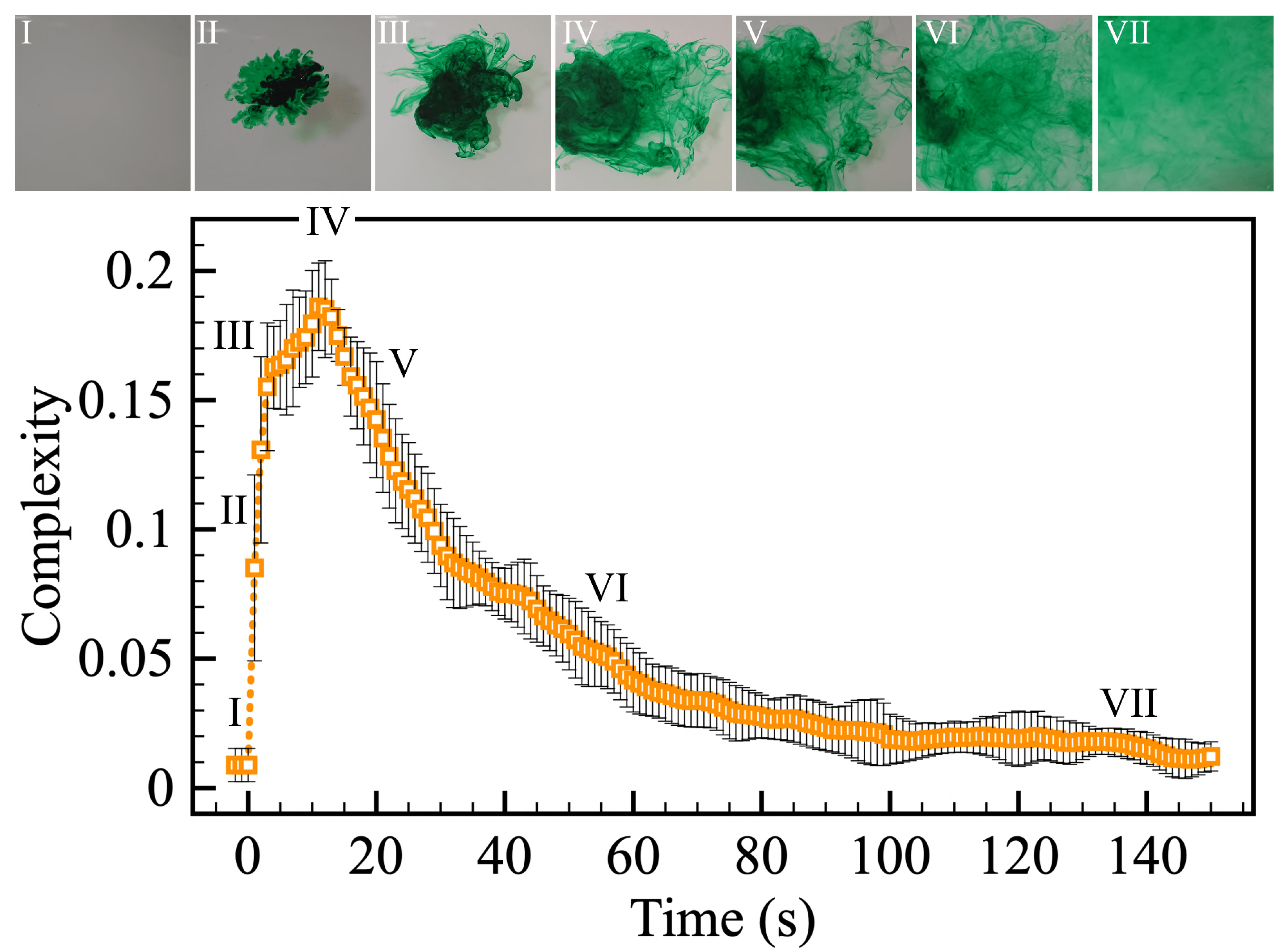}
\end{center}
\caption{The evolution of the complexity during the process of dissolving a food dye drop of 0.3 ml in water at 31$^{\circ}$C.}
\label{fig:green-drop-complexity}
\end{figure}
\subsection*{Complexity of time-dependent systems}
Finally, we would like to analyze how structural complexity evolves in time if entropy of the system is steadily increasing. The common wisdom is that computational complexity keeps increasing alongside the entropy, getting higher for more random states of the system. However, for structural complexity we should expect non-monotonous dependence on entropy.

To study this, we first move aside from the magnetic patterns case (we shall get back to it later) and take a look at the process of dissolving a dye drop in water. We put a $0.3$ ml drop of green dye in water at $31^\circ C$ and keep track of time evolution of the color spot. At every moment of time, state of the system is recorded as 2048$\times$2048 photo which is used to compute complexity of the apparent pattern. We have conducted the experiment six times and found that complexity as a function of time obeys quite a robust curve Fig. \ref{fig:green-drop-complexity}, with a quick increment stage followed by slow oscillatory fall-off at larger times.

\subsection*{Complexity of spin dynamics}

Since we have put the main focus on complexity of patterns emerging in magnetic systems, it is also natural to trace out time evolution of complexity in spin dynamics. As an example, we have consider processes occurring when a single magnetic skyrmion is perturbed by a picosecond magnetic field pulse. To simulate such skyrmion dynamics, one can use the following spin Hamiltonian~\cite{Blugel}
\begin{equation}\label{HamSD}
\begin{split}
H= - J\sum_{nn'}{\bf S}_n{\bf S}_{n'}-{\bf D}\sum_{nn'}[{\bf S}_n\times{\bf S}_{n'}] -K\sum_{n}({\bf S}_n^z)^2,
\end{split}
\end{equation}
 where $K$ is the strength of the uniaxial anisotropy in $z$ direction. We take into account only interactions between the nearest neighbours. The summation of inter-spin couplings runs twice over every pair. The Hamiltonian is defined on the 128$\times$128 square lattice. In our simulations, we used the following parameters: $J = 0.03676$ mRy, $|{\bf D}| = 0.008824$ mRy, and $K = 0.00735$ mRy. The Dzyaloshinskii-Moriya interaction vector is parallel to the lattice links.

Since we are interested in simulation of dynamical processes, to solve~\eqref{HamSD} we used the Landau-Lifshitz-Gilbert (LLG) equation as it is implemented in the Uppsala Atomistic Spin Dynamics (UppASD) software \cite{UppASD1, UppASD2}

\begin{equation}\label{LLG}
\begin{split}
\frac{d\textbf{S}_n}{dt}= -\frac{\gamma}{1+\alpha^2}\textbf{S}_n\times[-\frac{\partial H}{\partial\textbf{S}_n}+b_n(t)]-\\ -\frac{\gamma}{|\textbf{S}_n|}\frac{\alpha}{1+\alpha^2}\textbf{S}_n\times(\textbf{S}_n\times[-\frac{\partial H}{\partial\textbf{S}_n}+b_n(t)]),
\end{split}
\end{equation}
where $\gamma$ is the gyromagnetic ratio, $\alpha$ is the damping parameter, and $b_n(t)$ is a stochastic magnetic field with a Gaussian distribution arising from the thermal fluctuations. In this work, we take $\alpha = 0.36$.

To induce real-time dynamics in the system, we used a time-dependent magnetic field pulse defined by a Gaussian distribution as it was proposed in Ref.~\cite{Blugel},

\begin{figure}[!t]
\begin{center}
\includegraphics[width=\columnwidth]{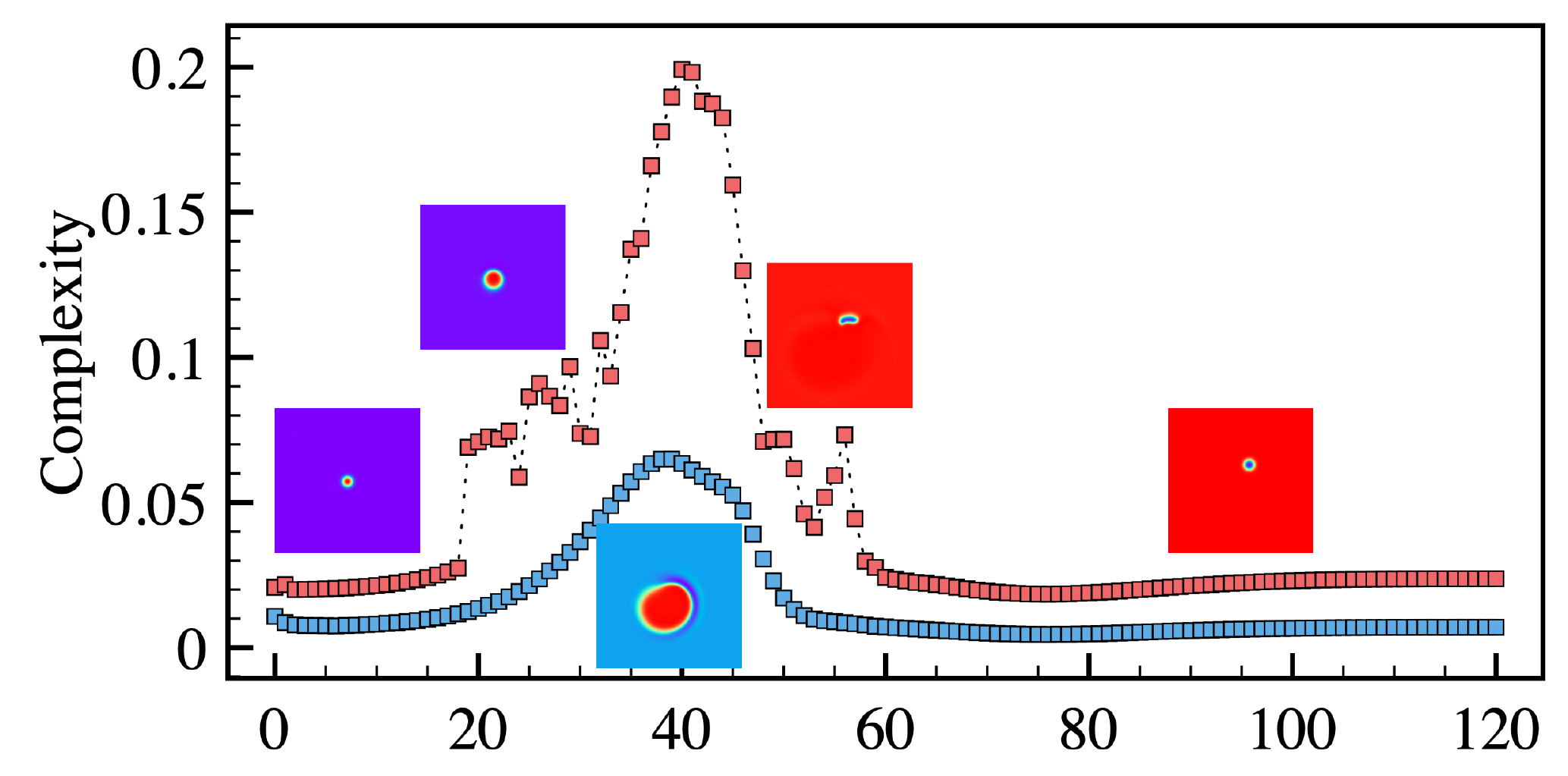}
\includegraphics[width=\columnwidth]{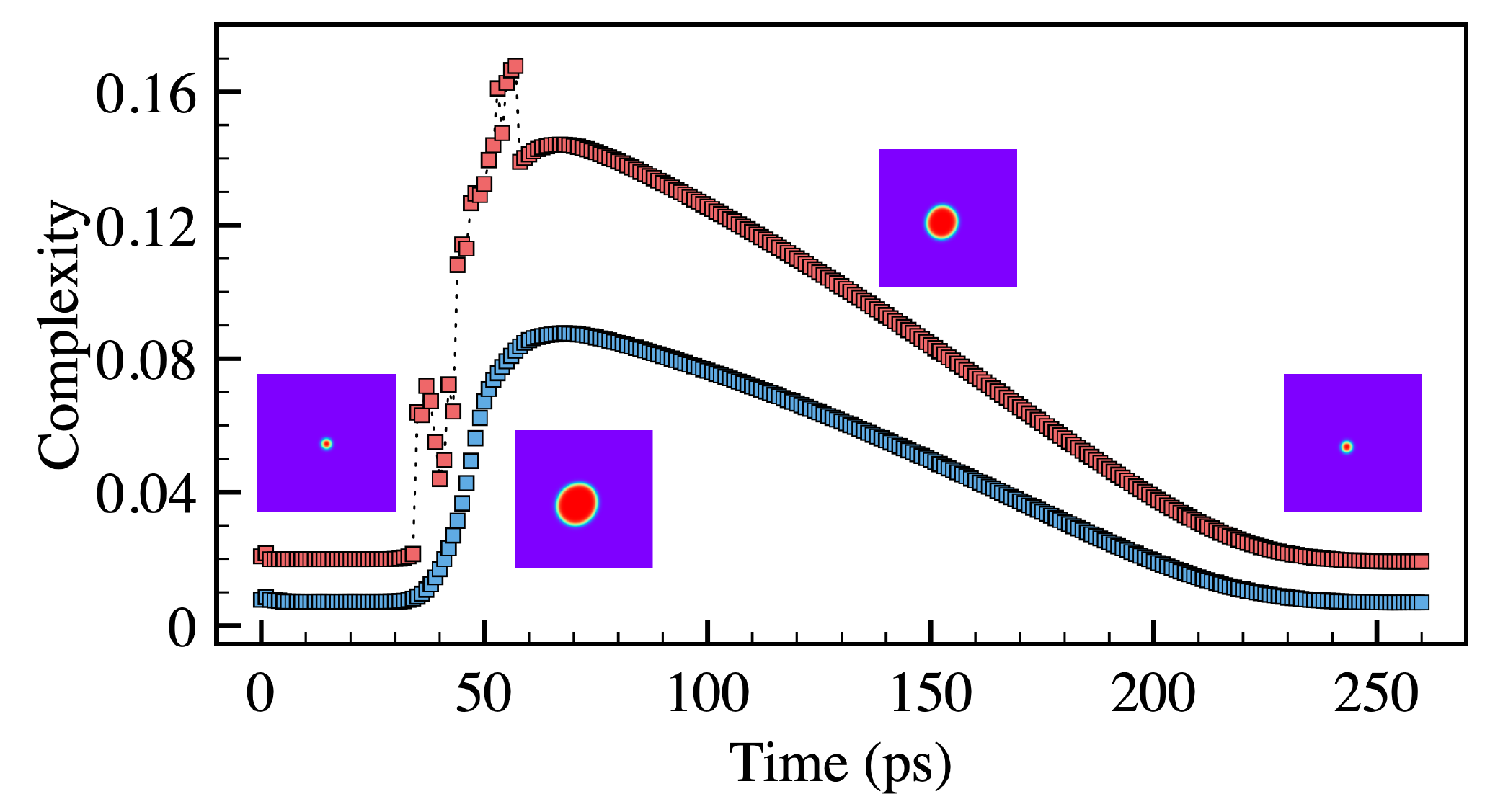}
\end{center}
\caption{The evolution of the complexity during the  switching (top panel) and breathing (bottom panel) processes generated with $t_{w}=28$ ps and $t_{w}=8$ ps, respectively. Red and blue squares represent the complexities calculated for 2048$\times$2048 images and 128$\times$128 Heisenberg spin arrays, respectively.}
\label{fig:skyrmion-dynamics}
\end{figure}

\begin{equation}\label{Gauss}
\begin{split}
\textbf{B}_p(t)= B_0exp\left(-\frac{(t-t_p)^2}{2t_w^2}\right)\textbf{e}_B,
\end{split}
\end{equation}
where $B_0$ is the amplitude of the magnetic field, $t_w$ is the Gaussian width, and $t_p$ is the time position of the pulse maximum. The real-space orientation of the magnetic pulse $\textbf{e}_B$ is described by the polar and azimuthal angles $\theta$ and $\varphi$. We used $B_0 = 2$ T, $t_p = 40$ ps, $\theta = 40^{\circ}$ and  $\varphi = 0^{\circ}$. The detailed phase diagram of processes realized in such system is given in our previous work~\cite{Deviatov}.

We have simulated so-called switching and breathing processes and computed the corresponding complexity dynamics using two different representations of data at each moment of time, - 128$\times$128 arrays of $z$-projections of the Heisenberg spins, and 2048$\times$2048 images visualizing state of the system. Simulations are performed with time step $\Delta t = 1$ ps. As can be seen from Fig.~\ref{fig:skyrmion-dynamics}, for both the switching and the breathing processes, the maximal value of ${\cal C}$ corresponds to the middle of the process, where the shape of the skyrmion is maximally perturbed. While the ``image'' representation leads to more noisy results due to the features that are artificially brought in by using higher resolution and a specific color scheme, it reflects the complexity dynamics of the ``true'' spin configuration pretty well, which indicates that the notion of complexity can be used in the cases when one has to deal with data in visualized form (images from different kinds of microscopes, diffraction patterns etc.). 

\begin{figure}[!h]
\begin{center}
\includegraphics[width=\columnwidth]{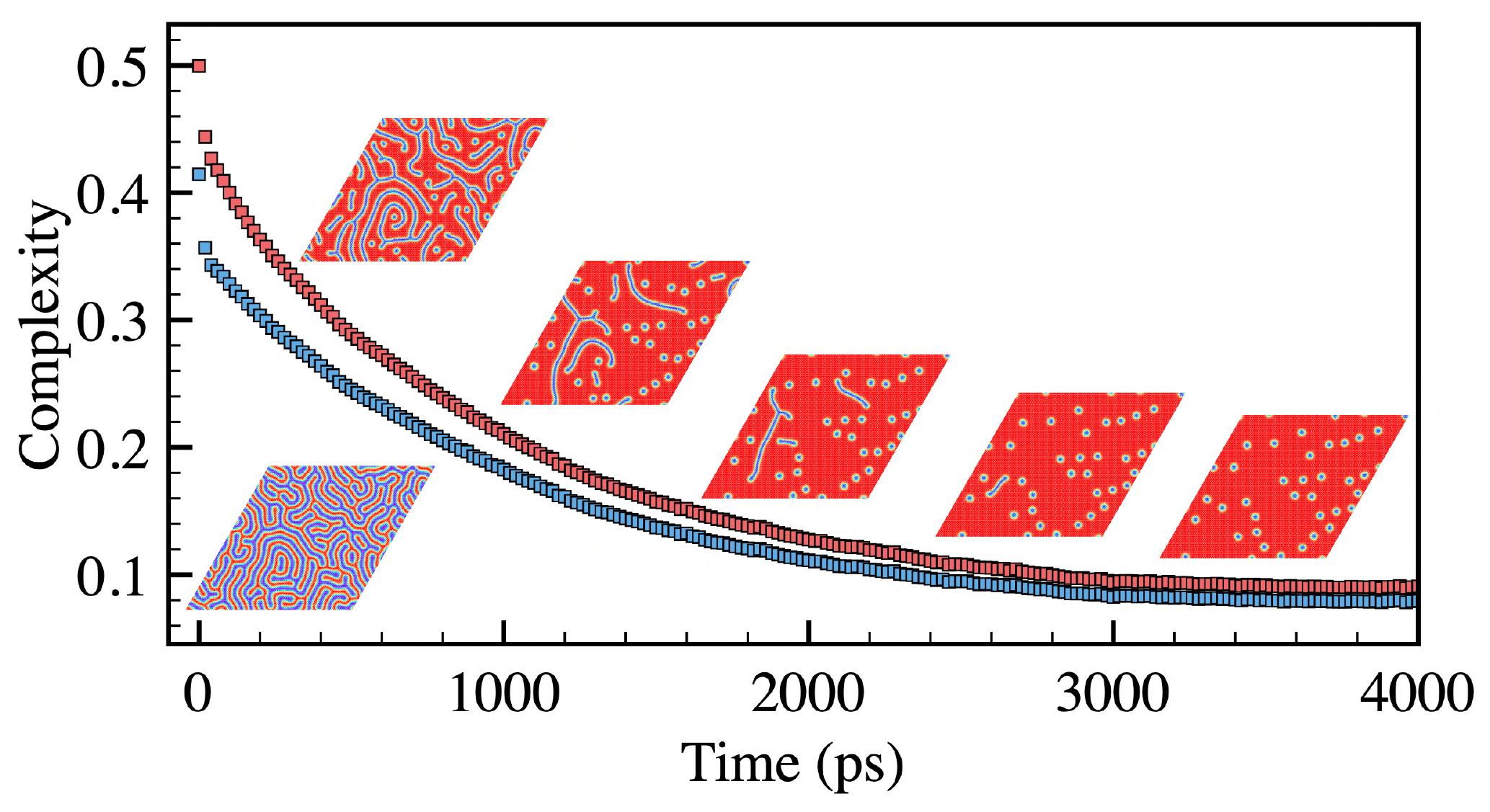}
\end{center}
\caption{The evolution of the complexity of the magnetic labyrinths on triangular lattice at $T = 0.02J$ after switching on the external magnetic field $B = 0.9J$ along $z$-axis. Here, $|{\bf D}|= J$. Red and blue squares correspond to the complexities calculated with $k \ge 0$ and $k \ge 1$, respectively.}
\label{fig:sp2sk}
\end{figure}

We have also studied the time evolution of magnetic labyrinths after switching on the external magnetic field. For this purpose, we define Hamiltonian~\eqref{Ham} on 128$\times$128 triangular lattice and choose the exchange parameters $|{\bf D}|= J$ (with $\bf D$ parallel to the links). Then we gradually cool down the system at zero magnetic field via Monte Carlo simulation and stabilize magnetic labyrinths. After that we apply $B = 0.9J$ and run spin dynamics simulations~\eqref{LLG} at $T = 0.02J$. As can be seen from Fig.~\ref{fig:sp2sk}, complexity slowly goes down as we move from spin spirals to disordered skyrmions.

\subsection*{Structural complexity as a Jackson integral}
While structural complexity defined by \eqref{eq:C_definition} is very easy to compute numerically, a better understanding of mathematical structures underlying this definition is highly desirable. To achieve that, we first consider the case when the pattern is defined as a one-dimensional discrete function on a lattice, $f(n):\mathbb{Z}\rightarrow\mathbb{R}$. Every step of the coarse-graining procedure assumes two operations: the first one is convolution of the pattern with some averaging function $R$ (usually called filter), and the second one is decimation that brings the smeared-out pattern to the smaller lattice size.
When computing the complexity, at every step first two patterns of the same size should be compared, - the original and the smeared-out ones, and only when this is done, the decimation is to be performed. 

In the Kadanoff decimation procedure, the lattice is divided in blocks of size $\Lambda$. Assuming that the filter has the same characteristic width, then smeared-out function $f_\text{conv}$ is defined by the formula:
\be\label{eq:fc}
f_\text{conv}(n)=(f*R)(\Lambda \Big[\frac{n}{\Lambda}\Big]),
\ee
where the square brackets denote the integer part, and the notation means that $f$ and $R$ first must be convolved, and the convolution is then evaluated at points $\Lambda \Big[\frac{n}{\Lambda}\Big]$ to assure that the resulting function assumes the same value at the points within the window of width $\Lambda$. 
In the momentum space, it can be approximated by a product
\be\label{eq:fc_mom}
\hat f_\text{conv}(k)=\hat f(k) \hat R(k).
\ee
In turn, convolution followed by decimation (a complete step of the renormalization procedure) can be defined as
\be
f_\text{dec}(n)=(f*R)(\Lambda n).
\ee
In the momentum space (this formula is exact for $\Lambda=2$):
\be
\hat f_\text{dec}(k)=\hat f(\Lambda^{-1} k)\hat R(\Lambda^{-1} k).
\ee
To make notations more concise, we denote function after $m$ complete steps of renormalization as $f_{m}$. Then, for its Fourier representation we get:
\be\label{eq:rec_rel}
\hat f_{m+1}(k)=\hat f_m(\Lambda^{-1} k)\hat R(\Lambda^{-1} k)
\ee
or
\be\label{eq:explicit_fm}
\hat f_{m}(k)=\hat f(\Lambda^{-m} k)\prod\limits^{m}_{j=1}\hat R(\Lambda^{-j} k).
\ee
Before, we defined partial contributions $C_{m}$ to the overall complexity as: 
\be\label{eq:C_m}
{\cal C}_{m}=\frac12\sum_{n}\left(f_{\text{conv},m}(n)-f_{m}(n)\right)^2.
\ee
Due to the Plancherel theorem, it has the same form in the momentum space:
\be
\sum_{n}\left(f_{\text{conv},m}(n)-f_{m}(n)\right)^2=\int dk\left(\hat{R}(k)\hat{f}_{m}(k)-\hat f_{m}(k)\right)^2. \nonumber
\ee
The overall complexity
\be
{\cal C} = \frac12 \sum\limits_{m=0}^{\infty} \int dk \left(\hat{R}(k)\hat{f}_{m}(k)-\hat f_{m}(k)\right)^2,
\ee
can be recast as:
\begin{gather}\label{eq:int_complex}
{\cal C}= \frac12 \int dk |\hat R(k)-1|^2 \sum\limits_{m=1}^{\infty}\hat{f}^2(\Lambda^{-m} k)\prod\limits^{m}_{j=1}\hat R^2(\Lambda^{-j} k) + \\
\frac12 \int dk |\hat R(k)-1|^2 \hat{f}^2(k), \nonumber
\end{gather}
where we explicitly extracted the contribution coming from dissimilarity between the original pattern and its first coarse-grained version\footnote{As was previously discussed, this part tends to accumulate random featureless complexity, and it is usually wise to ignore it.}.
Here we assume that the renormalization procedure can be formally conducted for an infinite number of steps. For any realistic finite-size pattern, it means that ${\cal C}_m$ contributions coming from large enough $m$ are simply zero, and there is no convergence issue.

To proceed further, it is handy to consider some typical example of the filter. By choosing $\hat R(k)$ to be a step filter with characteristic width $1/\Lambda$ in the momentum space which corresponds to the Wilson renormalization group:
\be
\hat R(k)=\Lambda\theta(-k+\frac{1}{2\Lambda})\theta(k+\frac{1}{2\Lambda})
\ee
which obeys
\be
\prod\limits_{j=1}^m \hat R(\Lambda^{-j}k)=
\Lambda^{m-1} \hat R(\Lambda^{-1} k), \label{eq:filter_product}
\ee
we arrive at the following expression for structural complexity \eqref{eq:int_complex}:
\begin{gather}
{\cal C}= \frac12 \int dk |\hat R(k) -1|^2 \hat R^2 (\Lambda^{-1}k) \sum_{m=1}^{\infty} \Lambda^{2m-2}  \hat f^2(\Lambda^{-m} k) +\\
\frac12 \int dk |\hat R(k) -1|^2\hat{f}^2(k). \nonumber
\end{gather}
By redefining $g(k)=\Lambda^{\frac{3m}{2}-1}\hat f(k)$, we obtain:
\begin{gather}
{\cal C}= \frac12 \int dk |\hat R(k) -1|^2 \hat R^2(\Lambda^{-1} k) \sum\limits_{m=1}^{\infty} \Lambda^{-m} g^2(\Lambda^{-m} k) +\\\frac12 \int dk |\hat R(k) -1|^2\hat{f}^2(k).
\nonumber
\end{gather}
This expression is quite remarkable as it can be recognized to contain a formal series defining Jackson integral from the quantum calculus \cite{QuantCalc}:
\be
\int f(k) d_q k = (1-q) k \sum\limits_{m=1}^\infty q^{m-1} f(q^{m-1} k).
\ee
By taking $q=\Lambda^{-1}$, the structural complexity can be represented as
\begin{gather} \label{eq:Jackson}
{\cal C}=\frac12 \int dk |\hat R(k) -1|^2 \hat{f}^2(k)+ \\ \frac{1}{2(\Lambda-1)}\int dk \frac{|\hat R(k) -1|^2 \hat R^2(\Lambda^{-1} k)}{k} \int g^2(\Lambda^{-1}k) d_{\Lambda^{-1}}k. \nonumber
\end{gather}
In deriving this equation, we relied on identity \eqref{eq:filter_product}. While it is exact for the step  filter in momentum space, it can be shown that for other types of filters it holds accurately making the Jackson integral representation of structural complexity an accurate approximation. For example, for the Gaussian filter
\be
\prod\limits_{j=1}^m \hat R(\Lambda^{-j}k)=\frac{\Lambda^{m/2}}{(2\pi)^{m/4}}\exp[-\frac{1}{4} \Lambda^2 k^2 \sum^{m}_{j=1}\Lambda^{-2j}],
\ee
and the following identity is pretty accurate\footnote{For $\Lambda=2$, the difference in $L^2$-norm between the two functions is about $14\%$.}:
\be
 \prod\limits_{j=1}^m \hat R(\Lambda^{-j}k)\simeq \frac{\Lambda^{m/2-1/2}}{(2\pi)^{m/4-1/4}} \hat R(\Lambda^{-1}k).
\ee
Defining $g(k) = \frac{\Lambda^{(2m-1)/2}}{(2\pi)^{(m-1)/4}} \hat f(k)$, one would again arrive at \eqref{eq:Jackson}.

\eqref{eq:Jackson} can be easily generalized onto higher dimensions as long as coarse-graining parameter $\Lambda$ is taken the same in all directions. E.g., for a 2d pattern $f(i,j): {\mathbb Z}^2 \rightarrow \mathbb R$:
\be\label{eq:1DRG}
f_{\Lambda}(i,j)=(f*R)(\Lambda i, \Lambda j),
\ee
and an identity similar to \eqref{eq:explicit_fm} is satisfied:
\be\label{eq:explicit_fm_2d}
\hat f_{m}(k_x,k_y)=\hat f(\Lambda^{-m} k_x, \Lambda^{-m} k_y)\Pi^{m}_{j=1}\hat R(\Lambda^{-j} k_x, \Lambda^{-j} k_y).
\ee
The rest is straightforward, and for the Wilsonian step filter we arrive at
\begin{gather}
{\cal C}= \frac12 \int d^2{\bf k} |\hat R({\bf k})-1|^2 \hat{f}^2({\bf k})+ \\ \frac12 \int d^2 {\bf k} \hat R^2(\Lambda^{-1} k_x, \Lambda^{-1} k_y)\cdot |\hat R({\bf k})-1|^2 \cdot \\ \sum\limits_{m=1}^{\infty} \Lambda^{-m} g^2(\Lambda^{-m} k_x, \Lambda^{-m} k_y), \nonumber
\end{gather}
where $g({\bf k})=\Lambda^{5m/2-2}\hat f({\bf k})$.

The appearance of Jackson integral related to quantum calculus in the context of structural complexity could be possibly understood in the connection with ultrametric analysis. Jackson integral can be viewed as an antiderivative analogue of the definite $p$-adic integral. In certain contexts, $p$-adic numbers are discussed as a natural candidate language for description of complex systems such as spin glasses \cite{Virasoro} or folding proteins because of the hierarchical structure of ${\mathbb Q}_p$ space \cite{Volovich, p-adic-image}. In most of the situations, the machinery of ultrametric analysis can be used only if the model of the phenomenon of interest is reformulated in $p$-adic space, which often puts strong limitations on what can actually be learned about the original real-space system. Contra to such cases, here, the corresponding ultrametric structures emerge naturally in the context of real-space patterns, which indicates that there could be a natural and universal connection between ultrametric analysis and complex systems\footnote{Speaking of possible relations between the theory of complex patterns and areas of modern mathematics, we would like to also refer to \cite{Kalinin}, where connections between self-organized criticality and tropical geometry were revealed.}.
\subsection*{Comparison with other approaches} 
\begin{figure}[!h]
\begin{center}
\includegraphics[width=0.8\columnwidth]{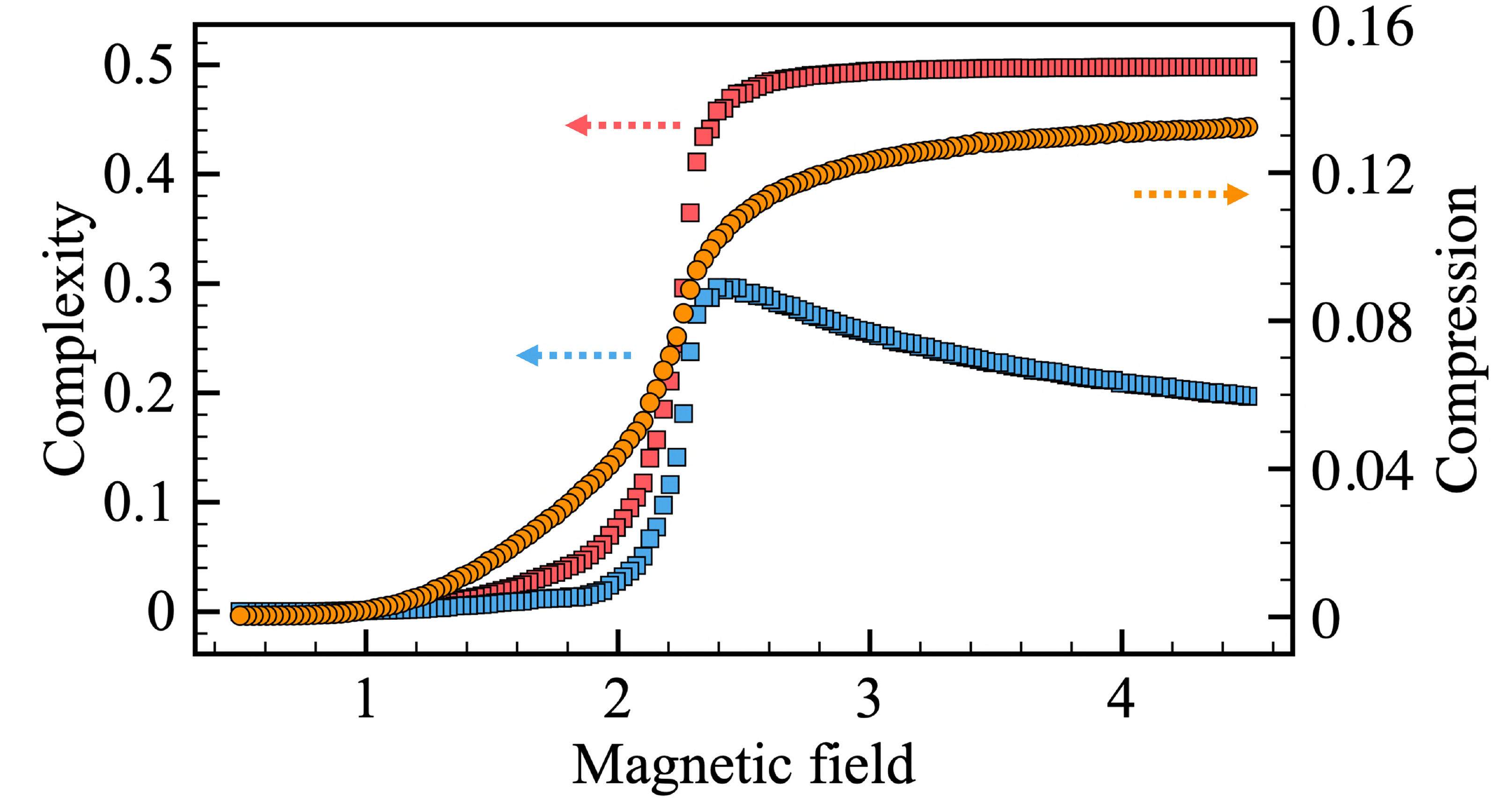}
\end{center}
\caption{Comparison between the structural complexity and compression ratio for two-dimensional Ising model simulations. Red and blue squares represent the complexities calculated with $k \ge 0$ and $k \ge 1$, respectively. Orange circles represent the compression rate.}
\label{is_vs_prx}
\end{figure}

\begin{figure}[!h]
\begin{center}
\includegraphics[width=0.8\columnwidth]{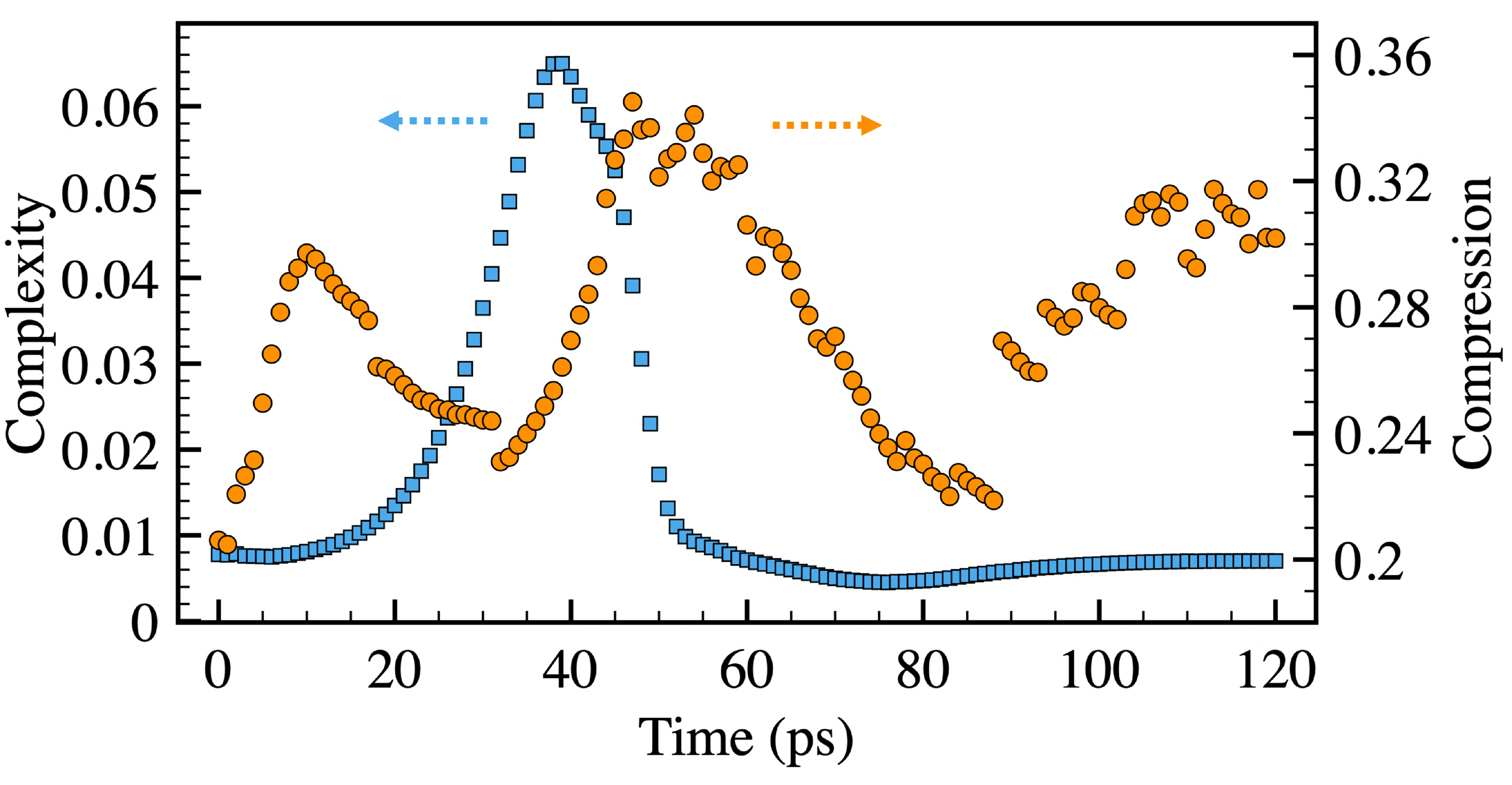}
\end{center}
\caption{Comparison between the structural complexity and compression rate for the switching process. Blue squares represent the complexity calculated for 128$\times$128 Heisenberg spins arrays, orange circles represent the compression rate.}
\label{sw_vs_prx}
\end{figure}

In the last few years, a set of methods for detecting phase transitions without knowing the order parameter, both in and out of equilibrium, based on loseless-compression algorithms have been suggested \cite{compress, compress1, compress2, compress3, compress4, compress5}. The main idea behind them is quite simple and transparent: the more ordered a system is, the shorter the description required to specify a typical state and the smaller the size ratio of the compressed and the original data files are. The compression rate can be viewed as another measure of complexity, and thus it is important to compare the notion of structural complexity proposed in this paper with the one which is already in use. We compare them by considering two particular examples, - the ferromagnetic-paramagnetic phase transition in the 2D Ising model, and the real-time dynamics of a single quenched skyrmion in the DM ferromagnet. To obtain the compression rates, we use the standard zip compressor (www.7-zip.org/download.html), and, in accordance with the prescription of \cite{compress}, represent the magnetic configurations as strings of numbers. One can see that both approaches give consistent results for the ferromagnetic-paramagnetic transition and allow to accurately estimate the critical temperature, Fig.~\ref{is_vs_prx}. At the same time, for the skyrmion switching, time evolution of the compression rate suffers from artifacts and does not really reflect how the process actually goes, Fig.~\ref{sw_vs_prx}. We tend to relate it to the fact that the Ising model state can be represented as a set of binaries (``up-down'' $\sim$ ``0-1''), and the symbolic representation does not introduce any artifacts. At the same time, when projection of a spin takes real values in $[-1,1]$, two very similar neighboring components can have symbolic representations that appear very different from the point of view of the compressor vocabulary, e.g., $s_z=0.593$ and $s_z=0.612$, and would cause a large undesired contribution to the compression complexity measure.

Another peculiarity of our method is that by construction it allows to analyze contributions of different characteristic scales to the resulting complexity, which is not possible within the compression method.

\subsection*{Conclusions}
In this paper, we have introduced a quantitative definition of effective complexity based on interscale dissimilarities of a system of interest. The system is assumed to be the more complex, the more distinctive features of different characteristic scales it has.

We exemplified this approach by computing complexity of certain $2d$ and $3d$ spatial structures, but it can be straightforwardly generalized onto any case that allows to define a coarse-graining protocol. 
Being a new easily computable measure, it might help to reveal some novel features of complex systems and processes. To conclude the paper, we shall outline our vision of possible applications of the suggested definition in different domains of science.

The most obvious and straightforward direction is to proceed further along the line of studying patterns emerging in classical solid state systems. Many types of experiments such as scanning tunneling microscopy, X-ray diffraction at synchrotrons and free-electron lasers, or neutron scattering experiments, produce visual patterns as an outcome of measurement. When the produced data forms a large set of images that are difficult to interpret (for example, if a previously unknown phase or structure has been observed \cite{Khaj-spinglass}), to digest relevant information might be a non-trivial task. In such cases, the concept of multi-scale complexity can be employed to detect novel phase transitions on the fly using raw experimental data.

In quantum science, the suggested concept can be utilized to define effective complexity of many-body wave functions that would complement the existent notions of computational circuit complexity \cite{Jefferson}. The usual types of metric on the Hilbert space are fidelity-based, meaning that two states are considered to be distinct if their overlap is low. However, in a many-body system, especially given the limited resolution of measurement device, it could be possible that two states with low overlap are in fact {\it operationally} similar, i.e. they are rather close in their experimentally accessible properties \cite{QiFock}. To formalize similarity/dissimilarity of this kind, the notion of structural complexity can be useful, as one can impose that two quantum states are close to each other, if their difference has low structural complexity. To accurately develop mathematical formalism of this kind requires additional effort (and we are planning to address this problem in a subsequent paper), but in the case of success, it could have a certain impact on the domain of quantum tomography, where one solves the task of accurately reconstructing quantum state from a limited number of low-order measurements \cite{Qtom}.

In the context of machine learning-based image recognition, the interscale complexity distribution could be used as an additional source of information about the features learned by neural networks. For example, an important problem in this field is how to recognize adversarial attacks, - {\it almost} invisible deformations of an image that make a trained neural network fail to properly identify which class it belongs to \cite{adversarial}. At this point, one cannot make a strong statement, but given the structure of deformations introduced by a typical adversarial attack, it is anticipated that such an attack would skew the complexity distribution of the image by increasing partial complexities at the microscopic scales, and the resulting anomaly can be identified at the preprocessing stage before feeding the image to the network.

Possible applications to biology are probably the most far-reaching goal, and it seems a bit speculative to discuss them in detail at this stage. Nevertheless, one can think of studying how complexity of genomic sequences evolves along different branches of the phylogenetic tree, and see whether major evolutionary transitions can be quantified in this way \cite{Koonin}. From a more general perspective, frustrations and competition between different levels of organization were claimed to be the most general moving force in appearance and development of biological complexity \cite{WKK}, which emphasizes the crucial role of {\it dis}similarity between representations of the system at different spacial and temporal scales. In these cases, the coarse-graining procedure required to compute structural complexity can be defined either on symbolic sequences of genomes or on visual representations of evolving biological structures (organelles or cells).

By no means, does our study give an exhaustive answer to the problem of quantifying effective complexity. After all, it is quite unlikely, that a unique universal definition should exist. Further studies are required to demonstrate how really useful the suggested measure is, but it is already clear that a number of research lines can be initiated on the basis of this approach.
\vskip 20pt

\acknow{We thank Yuri Bakhtin, Victor Kleptsyn, Eugene Koonin, Denis Kosygin, Slava Rychkov, Stanislav Smirnov, and Tom Westerhout for useful discussions, and Elena Mazurenko for technical assistance in conducting food dye experiments. The work of V.V.M., A.A.B., and I.A.I. was supported by the Russian Science Foundation, Grant No. 18-12-00185. M.I.K. acknowledges support by  Nederlandse Organisatie voor Wetenschappelijk Onderzoek (NWO) via Spinoza Prize.  A.A.I. acknowledges financial support from Dutch Science Foundation NWO/FOM under Grant No. 16PR1024. This work was partially supported by Knut and Alice Wallenberg Foundation through Grant No. 2018.0060. }

\showacknow 
\bibliography{pnas-sample}

\begin{thebibliography}{99}

\bibitem{adami}  C.~Adami, ``What is complexity?'', Bioessays {\bf 24}, 1085 (2002)

\bibitem{gellmann}  M. Gell-Mann, ``The Quark and the Jaguar: Adventures in the Simple and the Complex,'' St. Martin’s Griffin, New York, 1995

\bibitem{bakbook} P. Bak, ``How Nature Works: The Science of Self-Organized Criticality,'' Springer, New York, 1996

\bibitem{badii} R. Badii and A. Politi, ``Complexity. Hierarchical Structures and Scaling in Physics,'' Cambridge Univ. Press, Cambridge, 1997

\bibitem{KWK} M.~I.~Katsnelson, Y.~I.~Wolf, and E.~V.~Koonin, ``Towards physical principles of biological evolution,'' Phys. Scr. {\bf 93}, 043001 (2018)

\bibitem{Koonin} E.~V.~Koonin, ``The meaning of biological information,'' Phil. Trans. A {\bf 374}, 20150065 (2016)

\bibitem{WKK}  Y.~I.~Wolf, M.~I.~Katsnelson, and E.~V.~Koonin, ``Physical foundations of biological complexity,'' PNAS September 11, 2018 115 (37) E8678-E8687

\bibitem{MMC} S.~M.~Marshall, A.~R.~G.~Murray, and L.~Cronin, ``A probabilistic framework for identifying biosignatures using Pathway Complexity,'' Philos. Trans. R. Soc. A 375, 20160342 (2017) 

\bibitem{GL} M.~Gell-Mann and S.~Lloyd, ``Information measures, effective complexity, and total information,'' Complexity {\bf 2}: 44-52 (1996)

\bibitem{CarrNetwork}
M.~A.~Valdez, D.~Jaschke, D.~L.~Vargas, and L.~D.~Carr, ``Quantifying complexity in quantum phase transitions via mutual information complex networks,''  Phys. Rev. Lett. {\bf 119} (2017) 225301

\bibitem{Giulio}G.~Tononi, G.~M.~Edelman, and O.~Sporns, ``Complexity and coherency: integrating information in the brain,'' Trends in cognitive sciences 2.12 (1998): 474-484

\bibitem{DeGiuli}E. DeGiuli, ``Random language model,'' Phys. Rev. Lett. {\bf 122} (2019) 128301

\bibitem{beauty} S.~Lakhal, A.~Darmon, J.-P.~Bouchaud, M.~Benzaquen, Beauty and structural complexity, arXiv:1910.06088 (14 October 2019)

\bibitem{lloyd} S.~Lloyd, ``Measures of complexity: a nonexhaustive list,'' IEEE Control Systems Magazine, 21(4), 7-8 (2001)

\bibitem{Kolmogorov} A.~N.~Kolmogorov, ``Three approaches to the quantitative definition of information,'' Probl. Peredachi Inf., {\bf 1:1} (1965), 3–11

\bibitem{bak1} P.~Bak, C.~Tang, and K.~Wiesenfeld, ``Self-organized criticality: an explanation of the $1/f$ noise,'' Phys. Rev. Lett. {\bf 59} (1987) 381

\bibitem{bak2} P.~Bak and K.~Sneppen, ``Punctuated equilibrium and criticality in a simple model of evolution,'' Phys. Rev. Lett. {\bf 71} (1993) 4083

\bibitem{bak3} S.~Maslov, M.~Paczuski, and P.~Bak, ``Avalanches and $1/f$ noise in evolution and growth models,'' Phys. Rev. Lett. {\bf 73} (1994) 2162

\bibitem{geology} A.~Sornette and D.~Sornette, ``Self-Organized Criticality and earthquakes,'' Eur. Phys. Lett., {\bf 9} (1989) 197

\bibitem{wars} D.~C.~Roberts and D.~L.~Turcotte, ``Fractality and self-organized criticality of wars,'' Fractals, {\bf 6} (1998) 351

\bibitem{biol} N.~Vandewalle and M.~Ausloos, ``Self-organized criticality in phylogenetic-like tree growths,'' Journal de Physique I, 5(8) (1995) 1011-1025

\bibitem{RGK} I.~K.~Razumov, Yu.~N.~Gornostyrev, and M.~I.~Katsnelson, ``Autocatalytic mechanism of pearlite transformation in steel,'' Phys. Rev. Appl. {\bf 7} (2017) 014002

\bibitem{NonScaleFree} A.~D.~Broido, A.~Clauset, ``Scale-free networks are rare,'' Nat. Comm. 10, 1017 (2019)

\bibitem{Bennett} C.~H.~Bennett, ``Logical depth and physical complexity,'' in R.~Herken (ed.), The universal Turing machine, a half century survey, Oxford University Press, pp. 227-257 (1988)

\bibitem{Crutchfield}J.~P.~Crutchfield and K.~Young, ``Inferring statistical complexity,'' Phys. Rev. Lett. {\bf 63} (1989) 105

\bibitem{Wolpert1} D.~H.~Wolpert and W.~Macready, ``Using self‐dissimilarity to quantify complexity,'' Complexity {\bf 12}, pp.77-85 (2007)

\bibitem{Wolpert2}D.~H.~Wolpert and W.~Macready, ``Self-dissimilarity: An empirical measure of complexity,'' Sante Fe Institute, Santa Fe, NM, Working Paper (1997): 97-12

\bibitem{Lloyd-depth}S.~Lloyd and H.~Pagels, ``Complexity as thermodynamic depth,'' Annals of physics {\bf 188} (1988): 186-213

\bibitem{Sinelnikova} A.~Sinelnikova, A.~J.~Niemi, J.~Nilsson, and M.~Ulybyshev, ``Multiple scales and phases in discrete chains with application to folded proteins,'' Phys. Rev. E, {\bf 97} (2018) 052107

\bibitem{Susskind:2014}
  L.~Susskind,
  ``Computational complexity and black hole horizons,''
  Fortsch.\ Phys.\  {\bf 64}, 24 (2016),
    arXiv:1403.5695

\bibitem{Brown:2015}
  A.~R.~Brown, D.~A.~Roberts, L.~Susskind, B.~Swingle and Y.~Zhao,
  ``Holographic complexity equals bulk action?'',
  Phys.\ Rev.\ Lett.\  {\bf 116} (2016) 191301,
 arXiv:1509.07876
 
 \bibitem{Bagrov_complexity}D.~S.~Ageev, I.~Y.~Aref'eva, A.~A.~Bagrov and M.~I.~Katsnelson,
  ``Holographic local quench and effective complexity,''
  JHEP {\bf 1808} (2018) 071, 
  arXiv:1803.11162
  
\bibitem{Blugel}
C.~Heo, N.~S.~Kiselev, A.~K.~Nandy, S.~Bl\"ugel, T.~Rasing, ``Switching of chiral magnetic skyrmions by picosecond magnetic field pulses via transient topological states,'' Scientific Reports, \textbf{6}, (2016) 27146

\bibitem{pexels}
The photos and images used in this work were taken from www.pexels.com

\bibitem{Bialek1}
D.~L.~Ruderman and W.~Bialek, ``Statistics of natural images: Scaling in the woods,''
Phys. Rev. Lett. {\bf 73} (1994) 814

\bibitem{Bialek2}
G.J. Stephens, T. Mora, G. Tkačik, and W. Bialek, ``Statistical Thermodynamics of Natural Images,'' Phys. Rev. Lett. {\bf 110}, (2013) 018701 

\bibitem{Danilov} A.~A.~Bagrov, M.~Danilov, S.~Brener, M.~Harland, A.~I.~Lichtenstein, M.~I.~Katsnelson, ``Detecting quantum critical points in the $t$-$t'$ Fermi-Hubbard model via complex network theory,'' arXiv:1904.11463

\bibitem{Melko}
J.~Carrasquilla and R.~G.~Melko, ``Machine learning phases of matter,'' Nat. Phys. 13, 431 (2017)

\bibitem{Iakovlev1}
I. A. Iakovlev, O. M. Sotnikov, and V. V. Mazurenko, ``Supervised learning approach for recognizing magnetic skyrmion phases,'' Phys. Rev. B {\bf 98} (2018) 174411

\bibitem{Onsager}L.~Onsager, ``Crystal statistics. I. A two-dimensional model with an order-disorder transition,'' Phys. Rev. {\bf 65} (1944) 117

\bibitem{Fisher}
M.~E.~Fisher, ``The theory of equilibrium critical phenomena,'' Rep. Prog. Phys. {\bf 30} (1967) 615-730

\bibitem{MC}
A.~Sonsin, M.~Cortes, D.~R.~Nunes, J.~V.~Gomes, R.~S.~Costa, ``Computational analysis of 3D Ising model using Metropolis algorithms,'' Journal of Physics: Conference Series, 630 (2015) 

\bibitem{Dzyaloshinskii}I.~Dzyaloshinsky, ``A thermodynamic theory of ``weak'' ferromagnetism of antiferromagnetics,'' J. Phys. Chem. Solids {\bf 4} (1958) 241

\bibitem{Moriya}T.~Moriya, ``Anisotropic Superexchange Interaction and Weak Ferromagnetism,'' Phys. Rev. {\bf 120} (1960) 91

\bibitem{Iakovlev2}
I. A. Iakovlev, O. M. Sotnikov, and V. V. Mazurenko, ``Bimeron nanoconfined design,'' Phys. Rev. B {\bf 97} (2018) 184415

\bibitem{UppASD1}
B. Skubic, J. Hellsvik, L. Nordstr\"om, and O. Eriksson, ``A method for atomistic spin dynamics simulations: implementation and examples,'' J. Phys: Cond. Matt. \textbf{20}, (2008) 315203 

\bibitem{UppASD2}
O. Eriksson, A. Bergman, L. Bergqvist, and J. Hellsvik,
``Atomistic Spin Dynamics: Foundations and Applications,''
Oxford University Press, 2017

\bibitem{Deviatov}
A. Y. Deviatov, I. A. Iakovlev, and V. V. Mazurenko, ``Recurrent network classifier for ultrafast skyrmion dynamics,'' Phys. Rev. Applied {\bf 12} (2019) 054026

\bibitem{QuantCalc} V.~Kac, P.~Cheung, ``Quantum Calculus'', Universitext, Springer-Verlag, 2002

\bibitem{Virasoro}
R.~Rammal, G.~Toulouse, and M.~A.~Virasoro, ``Ultrametricity for physicists,'' Rev.~Mod.~Phys. {\bf 58} (1986) 765
\bibitem{Volovich} V.~S.~Vladimirov, I.~V.~Volovich, E.~I.~Zelenov, ``$p$-adic Analysis and Mathematical Physics'', World Scientific 1994

\bibitem{p-adic-image} P.~E.~Bradley, ``From image processing to topological modelling with p-adic numbers,'' P-Adic Numbers, Ultrametric Analysis, and Applications, 2(4), 293-304 (2010)

\bibitem{Kalinin} N.~Kalinin, A.~Guzmán-Sáenz, Y.~Prieto, M.~Shkolnikov, V.~Kalinina, and E.~Lupercio, ``Self-organized criticality and pattern emergence through the lens of tropical geometry,'' PNAS August 28, 2018 115 (35) E8135-E8142

\bibitem{compress} D. Sheinwald, A. Lempel, and J. Ziv, ``Two-Dimensional Encoding by
Finite-State Encoders,'' IEEE Trans. Commun. \textbf{38}, (1990) 341

\bibitem{compress1} Stefano Martiniani, Paul M. Chaikin, and Dov Levine, ``Quantifying
Hidden Order out of Equilibrium,'' Phys. Rev. X \textbf{9}, (2019) 011031

\bibitem{compress2} V.~Cortez, G.~Saravia, E.~E.~Vogel, ``Phase diagram and reentrance for the 3D Edwards-Anderson model using information theory,'' Journal of Magnetism and Magnetic Materials \textbf{372} (2014) 173-180

\bibitem{compress3} E.~E. Vogel, G.~Saravia, L.~V.~Cortez, ``Data compressor designed to improve recognition of magnetic phases,'' Physica A \textbf{391} (2012) 1591-1601

\bibitem{compress4} O. Melchert and A. K. Hartmann, ``Analysis of the phase transition in the two-dimensional Ising ferromagnet using a Lempel-Zivstring-parsing scheme and black-box data-compression utilities,'' Phys. Rev. E \textbf{91}, (2015) 023306

\bibitem{compress5} R.~Avinery, M.~Kornreich, and R.~Beck, ``Universal and Accessible
Entropy Estimation Using a Compression Algorithm,'' Phys. Rev. Lett. \textbf{123} (2019) 178102 

\bibitem{Khaj-spinglass} 
U. Kamber, A. Bergman, A. Eich, D. Iusan, M. Steinbrecher, N. Hauptmann, L. Nordstr\"om, M. I. Katsnelson, D. Wegner, O. Eriksson, and A. A. Khajetoorians, ``Self-induced spin glass state in elemental and crystalline neodymium,'' Science {\bf 368} (2020) 6757 

\bibitem{Jefferson} 
R.~A.~Jefferson, R.~C.~Myers, ``Circuit complexity in quantum field theory,'' JHEP 2017 (10), 107

\bibitem{QiFock} P.~Hosur and X.-L.~Qi, ``Characterizing eigenstate thermalization via measures in the Fock space of operators,'' Phys. Rev. E {\bf 93} (2016) 042138

\bibitem{Qtom} M.~Paris, J.~\v{R}eh\'{a}\v{c}ek (Eds.), ``Quantum state estimation,'' Lect. Notes Phys., 649, Springer, Berlin Heidelberg 2004

\bibitem{adversarial} A.~Madry, A.~Makelov, L.~Schmidt, D.~Tsipras, A.~Vladu, ``Towards Deep Learning Models Resistant to Adversarial Attacks,'', ICLR 2018 contribution, arXiv:1706.06083

\end{thebibliography}

\end{document}